\documentclass[12pt,eqno,epsf]{article}
\usepackage{amsmath,amssymb,graphicx,slashbox}
\numberwithin{equation}{section}

\def\mydate{May 6, 2016}
\def\ignore#1{{}}

\newcounter{sxn}

\newcounter{axn}

\date{}

\newdimen\mybaselineskip
\renewcommand{\thefootnote}{\arabic{footnote}}

\newcommand{\beeq}{\begin{equation}}
\newcommand{\eneq}{\end{equation}}
\newcommand{\beqn}{\begin{eqnarray}}
\newcommand{\eeqn}{\end{eqnarray}}

\newcommand{\alp}{\alpha}
\newcommand{\bt}{\beta}
\newcommand{\gm}{\gamma}
\newcommand{\Gm}{\Gamma}
\newcommand{\dlt}{\delta}

\newcommand{\ep}{\epsilon}
\newcommand{\zt}{\zeta}

\newcommand{\vth}{\vartheta}
\newcommand{\kp}{\kappa}
\newcommand{\lmd}{\lambda}
\newcommand{\Lmd}{\Lambda}
\newcommand{\sgm}{\sigma}

\newcommand{\vph}{\varphi}
\newcommand{\omg}{\omega}
\newcommand{\Omg}{\Omega}

\newcommand{\be}{\begin{equation}}
\newcommand{\ee}{\end{equation}}
\newcommand{\bea}{\begin{eqnarray}}
\newcommand{\eea}{\end{eqnarray}}
\newcommand{\eql}{\!\!\!&=\!\!\!&}

\newcommand{\defa}{\!\!\!&\equiv\!\!\!&}

\newcommand{\exch}{\leftrightarrow}

\newcommand{\tl}[1]{\tilde{#1}}
\newcommand{\bdm}[1]{{\mbox{\boldmath $#1$}}}
\newcommand{\sbdm}[1]{{\mbox{\scriptsize \boldmath $#1$}}}
\newcommand{\tr}{{\rm Tr}}

\newcommand{\diag}{{\rm diag}}
\newcommand{\der}{\partial}
\newcommand{\dr}{\!\!d}
\newcommand{\hc}{{\rm h.c.}}
\newcommand{\ie}{{i.e.}}
\newcommand{\id}{\mbox{\boldmath $1$}}
\newcommand{\sgn}{{\rm sgn}}

\newcommand{\vev}[1]{\langle #1 \rangle}

\newcommand{\brkt}[1]{\left( #1 \right)}
\newcommand{\brc}[1]{\left\{ #1 \right\}}
\newcommand{\sbk}[1]{\left[ #1 \right]}
\newcommand{\abs}[1]{\left| #1 \right|}

\renewcommand{\Re}{{\rm Re}\,}
\renewcommand{\Im}{{\rm Im}\,}

\newcommand{\thF}[2]{\vth\sbk{\begin{array}{c} #1 \\ #2 \end{array}}}

\newcommand{\cA}{{\cal A}}
\newcommand{\cB}{{\cal B}}
\newcommand{\cC}{{\cal C}}
\newcommand{\cD}{{\cal D}}
\newcommand{\cF}{{\cal F}}

\newcommand{\cL}{{\cal L}}
\newcommand{\cM}{{\cal M}}
\newcommand{\cN}{{\cal N}}
\newcommand{\cO}{{\cal O}}
\newcommand{\cP}{{\cal P}}
\newcommand{\cQ}{{\cal Q}}

\begin{document}
\thispagestyle{empty}

\baselineskip=12pt

{\small \noindent \mydate    
\hfill }

{\small \noindent \hfill KEK-TH-1888}

\baselineskip=35pt plus 1pt minus 1pt

\vskip 1.5cm

\begin{center}
{\Large\bf Yukawa couplings in 6D gauge-Higgs unification}\\
{\Large\bf on $\bdm{T^2/Z_N}$ with magnetic fluxes}

\vspace{1.5cm}
\baselineskip=20pt plus 1pt minus 1pt

\normalsize

{\bf Yoshio\ Matsumoto},${}^1\!${\def\thefootnote{\fnsymbol{footnote}}
\footnote[1]{\tt e-mail address: yoshio@post.kek.jp}} 
{\bf and 
Yutaka\ Sakamura}${}^{1,2}\!${\def\thefootnote{\fnsymbol{footnote}}
\footnote[2]{\tt e-mail address: sakamura@post.kek.jp}}

\vspace{.3cm}
${}^1${\small \it Department of Particles and Nuclear Physics, \\
SOKENDAI (The Graduate University for Advanced Studies), \\ 
Tsukuba, Ibaraki 305-0801, Japan} \\ \vspace{3mm}
${}^2${\small \it KEK Theory Center, Institute of Particle and Nuclear Studies, 
KEK, \\ Tsukuba, Ibaraki 305-0801, Japan} 
\end{center}

\vskip 1.0cm
\baselineskip=20pt plus 1pt minus 1pt

\begin{abstract}
We discuss the Yukawa couplings in 6D gauge-Higgs unification models on $T^2/Z_N$ 
in the presence of magnetic fluxes. 
We provide general formulae for them, and numerically evaluate their magnitude 
in a specific model on $T^2/Z_3$. 
Thanks to the nontrivial profiles of the zero-mode wave functions, 
the top quark Yukawa coupling can be reproduced without introducing 
a large representation of the gauge group for matter fields. 
However, it is difficult to realize small Yukawa couplings only by 
the magnetic fluxes and the Wilson-line phases because of 
the complicated structure of the mode functions on $T^2/Z_N$ ($N=3,4,6$). 
\end{abstract}


\newpage

\section{Introduction}
The gauge-Higgs unification (GHU)~\cite{Manton:1979kb}-\cite{Antoniadis:2001cv} 
is an interesting candidate 
for the new physics beyond the standard model. 
The Higgs fields are identified with extra-components of 
higher-dimensional gauge fields,  
and we do not need to introduce elementary scalar fields. 
The higher-dimensional gauge symmetry governs the Higgs and the Yukawa sectors. 
Namely, the gauge invariance prohibits the Higgs masses at tree-level,\footnote{
Six-dimensional (6D) models generically allow tadpole terms 
proportional to the field strength~$F_{45}$ at the orbifold fixed points. 
Such terms induce tree-level Higgs masses 
unless they are cancelled~\cite{Scrucca:2003ut}. 
} 
and the Yukawa couplings originate from the (higher-dimensional) gauge couplings. 
In particular, five-dimensional (5D) models have been extensively 
investigated~\cite{Agashe:2004rs}-\cite{Maru:2010ap} 
because they have the simplest extra-dimensional structure 
and the 5D gauge invariance protects the Higgs mass against large quantum corrections. 

Six-dimensional (6D) GHU models are also phenomenologically attractive because 
the existence of Higgs quartic couplings at tree-level makes 
a realization of the observed Higgs mass easier~\cite{Scrucca:2003ut}. 
In our previous work~\cite{Matsumoto:2014ila}, we investigated 6D GHU models 
on $T^2/Z_N$ orbifolds, and searched for possible gauge groups, 
orbifolds, and representations of the matter fermions 
by requiring the theory to have the custodial symmetry 
and realize the top quark mass. 
By employing the group theoretical analysis, 
we found that the minimal candidate is an U(4) gauge theory on $T^2/Z_3$ 
and the third-generation quarks are embedded into $\bdm{20'}$ of SU(4). 

6D models have another important feature. 
We can introduce magnetic fluxes that penetrate the compact space 
as a background. 
Such a background is phenomenologically interesting 
because it induces gauge symmetry breaking, chiral fermions 
in four-dimensional (4D) effective theories, and multiple zero-modes from a single bulk 
field~\cite{Bachas:1995ik}-\cite{Aldazabal:2000dg}. 
Besides, since the magnetic flux deforms the flat profile of zero-mode wave functions in 
the extra dimensions, it can control 4D effective Yukawa 
couplings~\cite{Cremades:2004wa}.


In this paper, we discuss the Yukawa couplings in 6D GHU models on $T^2/Z_N$ 
in the presence of background magnetic fluxes. 
As mentioned above, the Yukawa couplings originate from 
higher-dimensional gauge couplings. 
Hence, they become flavor-universal in a simple setup. 
In 5D models on $S^1/Z_2$, we can vary them by means of 
the bulk fermion masses that have kink profiles. 
Unfortunately, they cannot be extended to 6D models because we only have 
codimension 2 singularities on two-dimensional orbifolds. 
Instead, we can control them by the magnetic fluxes 
and the Wilson-line phases. 
Furthermore, the magnetic fluxes, which are quantized, can realize 
the generational structure of quarks and leptons. 
In Refs.~\cite{Abe:2008sx}-\cite{Abe:2015mua}, possibilities of reproducing 
the realistic Yukawa structure by magnetic fluxes are investigated 
in the context of ten-dimensional super Yang-Mills or superstring theories, 
and it is shown to be reproduced in some cases. 
Their success of the realization of the Yukawa hierarchy 
is supported by the following two points. 
One is that the gauge groups they considered are large and contain a lot of U(1) subgroups 
that have the magnetic fluxes,  
which means that there exist sufficient number of independent magnetic fluxes 
to control the Yukawa couplings. 
The other is that their models are compactified on $T^2$ or $T^2/Z_2$.\footnote{
In Ref.~\cite{Abe:2015yva}, the cases in which three generations are realized are 
discussed on $T^2/Z_N$ ($N=2,3,4,6$). 
However, the numerical evaluations of the Yukawa couplings are performed 
only on $T^2/Z_2$. 
} 
Hence the mode functions have simpler structures than those on $T^2/Z_N$ ($N=3,4,6$), 
and easier to control. 
However, these properties are not necessary conditions for the GHU models. 
In this paper, we discuss realization of the Yukawa hierarchy 
in smaller gauge groups, and especially focus on a U(3) model on $T^2/Z_3$ 
as a specific example. 


The paper is organized as follows. 
In the next section, we explain our setup and introduce the magnetic fluxes. 
In Sec.~\ref{md_fct}, we show explicit forms of the mode functions on $T^2$ and $T^2/Z_N$. 
In Sec.~\ref{Yukawa_cp}, we provide a formula for the Yukawa coupling constants, 
and evaluate their numerical values in a specific model. 
Sec.~\ref{summary} is devoted to the summary.

\section{Setup}
We consider a 6D gauge theory compactified on an orbifold~$T^2/Z_N$ ($N=2,3,4,6$). 
The gauge group is $G\times {\rm U(1)}_X$, where $G$ is 
a simple group that includes ${\rm SU(2)}_{\rm L}\times {\rm U(1)}_Z$.\footnote{
We do not consider the color group~${\rm SU(3)_C}$ 
since it is irrelevant to the discussion,  
and ${\rm U(1)}_X$ is introduced in order to adjust the Weinberg angle 
to the realistic value. 
}
The field content consists of the $G$ gauge field~$A_M$, 
the ${\rm U(1)}_X$ gauge field~$B_M$, 
where $M=0,1,\cdots,5$ is the 6D Lorentz index, 
and 6D Weyl fermions~$\Psi^f_{\chi_6}$ ($f=1,2,\cdots$), 
where $\chi_6=\pm$ denotes the 6D chirality. 
The 6D Lagrangian is
\bea
 \cL \eql -\frac{1}{4g_A^2}\tr\brkt{F^{MN}F_{MN}}
 -\frac{1}{4g_B^2}B^{MN}B_{MN}+\cL_{\rm gf}
 +\sum_fi\bar{\Psi}^f_{\chi_6}\Gm^M\cD_M\Psi^f_{\chi_6}, 
 \label{6DcL}
\eea
where $\cL_{\rm gf}$ denotes the gauge-fixing terms, $\Gm^M$ are 6D gamma matrices, 
and $g_A$ and $g_B$ are 
the 6D gauge coupling constants for $G$ and ${\rm U(1)}_X$, respectively. 
The field strengths and the covariant derivatives are defined as
\bea
 F_{MN} \defa \der_M A_N-\der_N A_M-i\sbk{A_M,A_N}, \nonumber\\
 B_{MN} \defa \der_M B_N-\der_N B_M, \nonumber\\
 \cD_M\Psi^f_{\chi_6} \defa 
 \brkt{\der_M-iA_M-iq_fB_M}\Psi^f_{\chi_6}, 
\eea
where $q_f$ is the ${\rm U(1)}_X$ charge of $\Psi^f_{\chi_6}$.

\subsection{Orbifold and boundary conditions}
%
For the coordinates of the extra dimensions, 
it is convenient to use a complex (dimensionless) 
coordinate~$z\equiv\frac{1}{2\pi R_1}(x^4+ix^5)$, 
where $R_1>0$ is one of the radii of $T^2$. 
Correspondingly, the extra dimensional components of the gauge fields 
are written as   
\be
 A_z = \pi R_1\brkt{A_4-iA_5}, \;\;\;\;\;
 B_z = \pi R_1\brkt{B_4-iB_5}.  
\ee
The orbifold~$T^2/Z_N$ is defined by identifying points in the extra space as 
\be
 z \sim \omg z+n_1+n_2\tau, \;\;\;\;\; (n_1,n_2\in\mathbb{Z})
\ee
where $\omg\equiv e^{2\pi i/N}$ and $\tau$ is a complex constant 
that satisfies $\Im\tau>0$. 
An arbitrary value of $\tau$ is allowed when $N=2$ 
while it must be equal to $\omg$ when $N=3,4,6$. 
The orbifold~$T^2/Z_N$ has the following fixed points 
in the fundamental domain~\cite{Kobayashi:1991rp}:
\be
 z = z_f \equiv \begin{cases} 0, \frac{1}{2}, \frac{\tau}{2}, \frac{1+\tau}{2}, 
 & (\mbox{on $T^2/Z_2$}) \\
 0, \frac{2+\tau}{3}, \frac{1+2\tau}{3}, & (\mbox{on $T^2/Z_3$}) \\
 0, \frac{1+\tau}{2}, & (\mbox{on $T^2/Z_4$}) \\
 0. & (\mbox{on $T^2/Z_6$}) \end{cases}
\ee
We can introduce 4D fields or interactions at these fixed points.  
Fields at equivalent points on $T^2/Z_N$ do not have to be equal 
as long as the Lagrangian is single-valued. 
The torus boundary conditions are expressed as
\bea
 A_M(x,z+s) \eql U_s(z)A_M(x,z)U_s^{-1}(z)+i(U_s\der_M U_s^{-1})(z), \nonumber\\
 B_M(x,z+s) \eql B_M(x,z)+\der_M\Lmd_s(z), \nonumber\\
 \Psi^f_{\chi_6}(x,z+s) \eql e^{iq_f\Lmd_s(z)}U_s(z)\Psi^f_{\chi_6}(x,z), \;\;\;\;\;
 \label{torusBC}
\eea
where $s=1,\tau$. 
Matrices~$U_s(z)\in G$ and real functions~$\Lmd_s(z)$ may depend on $z$. 
The orbifold boundary conditions are 
\bea
 A_\mu(x,\omg z) \eql P A_\mu(x,z) P^{-1}, \;\;\;\;\;
 A_z(x,\omg z) = \omg^{-1}PA_z(x,z)P^{-1}, \nonumber\\
 B_\mu(x,\omg z) \eql B_\mu(x,z), \;\;\;\;\;
 B_z(x,\omg z) = \omg^{-1}B_z(x,z), \nonumber\\
 \Psi^f_{\chi_6,\chi_4}(x,\omg z) \eql 
 \omg^{-\frac{\chi_4\chi_6}{2}}e^{i\vph_f}P\Psi^f_{\chi_6,\chi_4}(x,z), 
 \label{orbifoldBC}
\eea
where $\chi_4$ denotes the 4D chirality, $\vph_f$ and $P\in G$ are a real constant 
and a constant matrix, respectively. 

The $G$ gauge field is decomposed as
\be
 A_M = \sum_iC_M^iH_i+\sum_{\sbdm{\alp}}W_M^{\sbdm{\alp}}E_{\sbdm{\alp}}, 
\ee
where $\{H_i,E_{\sbdm{\alp}}\}$ are the generators of $G$ in the Cartan-Weyl basis, 
\ie, $H_i$ ($i=1,2,\cdots,{\rm Rank}\,G$) are the Cartan generators 
and $\bdm{\alp}$ runs over all the roots of $G$. 
The generators are normalized as $\tr(H_iH_j)=\dlt_{ij}$ 
and $\tr(E_{\sbdm{\alp}}E_{\sbdm{\bt}})=\dlt_{\sbdm{\alp},-\sbdm{\bt}}$. 
We can always choose the generators 
in such a way that $P$ in (\ref{orbifoldBC}) is expressed as
\be
 P = \exp\brkt{ip\cdot H}, 
\ee
where $p\cdot H\equiv \sum_ip^iH_i$ ($p^i$: real constants). 
Since (\ref{orbifoldBC}) is a $Z_N$-transformation, 
the following relations must hold: 
\bea
 e^{ip\cdot\sbdm{\alp}} \eql \exp\brkt{\frac{2n_{\sbdm{\alp}}\pi i}{N}}, \nonumber\\
 \omg^{-\frac{\chi_4\chi_6}{2}}e^{i\vph_f}e^{ip\cdot\sbdm{\mu}} 
 \eql \exp\brkt{\frac{2n_{\sbdm{\mu}f}^{\chi_4\chi_6}\pi i}{N}}, 
 \label{cstrtonvph_f}
\eea
where $n_{\sbdm{\alp}},n_{\sbdm{\mu}f}^{\chi_4\chi_6}\in\mathbb{Z}$.

\subsection{Magnetic fluxes}
We introduce the magnetic fluxes that penetrate $T^2/Z_N$ as a background. 
For simplicity, we assume that $W_M^{\sbdm{\alp}}$ do not have 
nonvanishing background and the background values of the field strengths are constants. 
Then nonvanishing constant fluxes are 
\bea
 \cC^i \defa \int_{T^2/Z_N}\dr x^4dx^5\;\vev{C_{45}^i} = \cA\vev{C_{45}^i} 
 = -\frac{2i\Im\tau}{N}\vev{C_{z\bar{z}}^i}, \nonumber\\
 \cB \defa \int_{T^2/Z_N}\dr x^4dx^5\;\vev{B_{45}} = \cA\vev{B_{45}} 
 = -\frac{2i\Im\tau}{N}\vev{B_{z\bar{z}}}, 
\eea
where $C_{z\bar{z}}^i=\der_z C_{\bar{z}}^i-\der_{\bar{z}}C_z^i$, 
$B_{z\bar{z}}=\der_z B_{\bar{z}}-\der_{\bar{z}}B_z$, 
and $\cA\equiv (2\pi R_1)^2\Im\tau/N$ is the area of the fundamental domain of $T^2/Z_N$. 
This indicates that 
the vector potentials~$C_z^i$ and $B_z$ have the following background values:
\be
 \vev{C_z^i} = -\frac{iN(\cC^i\bar{z}+\bar{c}^i)}{4\Im\tau}, \;\;\;\;\;
 \vev{B_z} = -\frac{iN(\cB\bar{z}+\bar{b})}{4\Im\tau}, 
 \label{bg:AB_z}
\ee
where $c^i$ and $b$ are complex constants, 
which correspond to the Wilson-line phases~\cite{Cremades:2004wa,Abe:2013bca}. 
From (\ref{bg:AB_z}), we identify $U_s(z)$ and $\Lmd_s(z)$ ($s=1,\tau$) 
in (\ref{torusBC}) as
\bea
 U_s(z) \eql \exp\brc{i\sum_i\brkt{\frac{N\cC^i\Im(\bar{s}z)}{2\Im\tau}
 +2\pi\alp_s^i}H_i}, \nonumber\\
 \Lmd_s(z) \eql \frac{N\cB\Im(\bar{s}z)}{2\Im\tau}+2\pi\bt_s, 
 \label{def:Uchi}
\eea
where $\alp_s^i$ and $\bt_s$ are real constants, which correspond to 
the Scherk-Schwarz (SS) phases~\cite{Cremades:2004wa,Abe:2013bca}. 
The magnetic fluxes~$\cC^i$ and $\cB$ are quantized as
\bea
 N\cC\cdot\bdm{\alp} \eql 2k_{\sbdm{\alp}}\pi, \nonumber\\
 N\brkt{\cC\cdot\bdm{\mu}+q_f\cB} \eql 2k_{\sbdm{\mu}f}\pi, 
 \label{quantize:CB}
\eea
where $\bdm{\alp}$ and $\bdm{\mu}$ are a root and a weight of $G$, 
and $k_{\sbdm{\alp}},k_{\sbdm{\mu}f}\in\mathbb{Z}$. 
The first and the second conditions originate from the requirement 
for the single-valuedness of $W_z^{\sbdm{\alp}}$ and $\Psi^f$ on $T^2/Z_N$, 
respectively. 
Using (\ref{quantize:CB}), the background gauge fields are expressed as
\bea
 \vev{C_z\cdot\bdm{\alp}} \eql 
 -\frac{k_{\sbdm{\alp}}\pi i(\bar{z}+\bar{\zeta}_{\sbdm{\alp}})}{2\Im\tau}, \nonumber\\
 \vev{C_z\cdot\bdm{\mu}+q_fB_z} \eql 
 -\frac{k_{\sbdm{\mu}f}\pi i(\bar{z}+\bar{\zeta}_{\sbdm{\mu}f})}{2\Im\tau}, 
\eea
where 
\be
 \zeta_{\sbdm{\alp}} \equiv \frac{c\cdot\bdm{\alp}}{\cC\cdot\bdm{\alp}}, \;\;\;\;\;
 \zeta_{\sbdm{\mu}f} \equiv \frac{c\cdot\bdm{\mu}+q_fb}{\cC\cdot\bdm{\mu}+q_f\cB}. 
 \label{def:zt}
\ee

We assume that the magnetic fluxes break $G$ 
to ${\rm SU(2)}_{\rm L}\times {\rm U(1)}_X\times {\rm U(1)}^{r-2}$ 
($r$: rank of $G$), 
and that ${\rm U(1)}_Z\times {\rm U(1)}_X$ is broken down to 
the hypercharge group~${\rm U(1)}_Y$ at one of the orbifold fixed points 
by some dynamics. 
The generators of the unbroken ${\rm SU(2)}_{\rm L}$ and ${\rm U(1)}_Z$ are 
expressed as
\be
 \brkt{T_{\rm L}^\pm,T_{\rm L}^3} = 
 \brkt{\frac{E_{\pm\sbdm{\alp}_{\rm L}}}{\abs{\bdm{\alp}_{\rm L}}},
 \frac{\bdm{\alp}_{\rm L}\cdot H}{\abs{\bdm{\alp}_{\rm L}}^2}}, \;\;\;\;\;
 \cQ_Z = \bdm{\eta}\cdot H, \label{def:alp_L}
\ee
where $\bdm{\alp}_{\rm L}$ is a root of ${\rm SU(2)}_{\rm L}\subset G$, 
and a constant real vector~$\bdm{\eta}$ satisfies 
$\bdm{\eta}\cdot\bdm{\alp}_{\rm L}=0$. 
Then the hypercharge~$Y$ is expressed in terms of $\cQ_Z$ and 
the ${\rm U(1)}_X$ generator~$\cQ_X$ as
\be
 Y = \cQ_Z+\cQ_X.  \label{def:Y}
\ee

\section{Mode functions} \label{md_fct}
In this section, we provide a brief review of the results in 
Refs.~\cite{Cremades:2004wa,Abe:2015yva,Abe:2008fi,Abe:2013bca,Abe:2014noa} in our notations, 
and show explicit forms of the mode functions on $T^2$ and $T^2/Z_N$.

\subsection{Kaluza-Klein mode expansion}
The 6D fields are expanded into the Kaluza-Klein (KK) modes as 
\bea
 C_\mu^i(x,z) \eql \frac{g_A}{\sqrt{2}\pi R_1}\sum_n f_n^i(z)C_\mu^{i(n)}(x), \;\;\;\;\;
 W_\mu^{\sbdm{\alp}}(x,z) 
 = \frac{g_A}{\sqrt{2}\pi R_1}\sum_nf_n^{\sbdm{\alp}}(z)W_\mu^{\sbdm{\alp}(n)}(x), \nonumber\\
 B_\mu(x,z) \eql \frac{g_B}{\sqrt{2}\pi R_1}\sum_n f_n^B(z) B_\mu^{(n)}(x), \nonumber\\
 C_z^i(x,z) \eql \vev{C_z^i}(z)+g_A\sum_n g_n^i(z)\vph_n^i(x), 
 \;\;\;\;\;
 W_z^{\sbdm{\alp}}(x,z) = g_A\sum_n 
 g_n^{\sbdm{\alp}}(z)\vph_n^{\sbdm{\alp}}(x), \nonumber\\
 B_z(x,z) \eql \vev{B_z}(z)+g_B\sum_n g_n^B(z)\vph_n^{B}(x), \nonumber\\
 \psi_{\pm}^f(x,z) \eql \frac{1}{\sqrt{2}\pi R_1}\sum_n\sum_{\sbdm{\mu}}
 h_{{\rm R}n}^{(\pm)\sbdm{\mu}f}(z)|\bdm{\mu}\rangle
 \psi_{\pm n}^{\sbdm{\mu}f}(x), \nonumber\\
 \bar{\lmd}_{\pm}^f(x,z) \eql \frac{1}{\sqrt{2}\pi R_1}\sum_n\sum_{\sbdm{\mu}}
 h_{{\rm L}n}^{(\pm)\sbdm{\mu}f}(z)|\bdm{\mu}\rangle
 \bar{\lmd}_{\pm n}^{\sbdm{\mu}f}(x), 
 \label{KKexpand}
\eea
where $|\bdm{\mu}\rangle$ is a vector in the $G$ representation space 
that corresponds to the weight~$\bdm{\mu}$. 
The fermion fields~$\psi_\pm^f$ and $\bar{\lmd}_\pm^f$ are 
the right- and the left-handed 2-component spinors defined as
\bea
 \Psi_+^f \eql \begin{pmatrix} \hat{\Psi}_+^f \\ \bdm{0}_4 \end{pmatrix}, \;\;\;\;\;
 \hat{\Psi}_+^f = \begin{pmatrix} \psi_{+\alp}^f \\ \bar{\lmd}_+^{f\dot{\alp}} \end{pmatrix}, 
 \nonumber\\
 \Psi_-^f \eql \begin{pmatrix} \bdm{0}_4 \\ \hat{\Psi}_-^f \end{pmatrix}, \;\;\;\;\;
 \hat{\Psi}_-^f = \begin{pmatrix} \psi_{-\alp}^f \\ \bar{\lmd}_-^{f\dot{\alp}} \end{pmatrix}. 
\eea
All the mode functions are defined to be dimensionless, 
and normalized as
\be
 \int_{T^2/Z_N}\dr zd\bar{z}\;F_n^*(z)F_m(z) = \dlt_{nm}, 
\ee
where $F_n(z)$ denotes the mode functions. 
The coefficients in the KK expansion are determined so that the 4D KK modes have 
canonically normalized kinetic terms.\footnote{
Note that $\int\dr x^4 dx^5 = 2(\pi R_1)^2\int\dr z d\bar{z}$. 
}

From (\ref{torusBC}) and (\ref{def:Uchi}), the mode functions should satisfy 
\bea
 f_n^i(z+s) \eql f_n^i(z), \;\;\;\;\;
 f_n^B(z+s) = f_n^B(z), \nonumber\\
 f_n^{\sbdm{\alp}}(z+s) \eql \exp\brc{\frac{k_{\sbdm{\alp}}\pi i}{\Im\tau}
 \Im(\bar{s}z)+2\pi i\phi_s^{\sbdm{\alp}}}f_n^{\sbdm{\alp}}(z), \nonumber\\
 g_n^i(z+s) \eql g_n^i(z), \;\;\;\;\;
 g_n^B(z+s) = g_n^B(z), \nonumber\\
 g_n^{\sbdm{\alp}}(z+s) \eql \exp\brc{\frac{k_{\sbdm{\alp}}\pi i}{\Im\tau}
 \Im(\bar{s}z)+2\pi i\phi_s^{\sbdm{\alp}}}g_n^{\sbdm{\alp}}(z), \nonumber\\
 h_{{\rm R}n}^{(\pm)\sbdm{\mu}f}(z+s) \eql 
 \exp\brc{\frac{k_{\sbdm{\mu}f}\pi i}{\Im\tau}\Im(\bar{s}z)+2\pi i\phi_s^{\sbdm{\mu}f}}
 h_{{\rm R}n}^{(\pm)\sbdm{\mu}f}(z), \nonumber\\
 h_{{\rm L}n}^{(\pm)\sbdm{\mu}f}(z+s) \eql 
 \exp\brc{\frac{k_{\sbdm{\mu}f}\pi i}{\Im\tau}\Im(\bar{s}z)+2\pi i\phi_s^{\sbdm{\mu}f}}
 h_{{\rm L}n}^{(\pm)\sbdm{\mu}f}(z), 
 \label{BC:md_eq:1}
\eea
where 
\be 
 \phi_s^{\sbdm{\alp}} \equiv \alp_s\cdot\bdm{\alp}, \;\;\;\;\;
 \phi_s^{\sbdm{\mu}f} \equiv \alp_s\cdot\bdm{\mu}+q_f\bt_s, 
\ee
and from (\ref{orbifoldBC}), they also satisfy 
\bea
 f_n^i(\omg z) \eql f_n^i(z), \;\;\;\;\;
 f_n^{\sbdm{\alp}}(\omg z) = e^{ip\cdot\sbdm{\alp}}f_n^{\sbdm{\alp}}(z), \;\;\;\;\;
 f_n^B(\omg z) = f_n^B(z), \nonumber\\
 g_n^i(\omg z) \eql \omg^{-1}g_n^i(z), \;\;\;\;\;
 g_n^{\sbdm{\alp}}(\omg z) = \omg^{-1}e^{ip\cdot\sbdm{\alp}}g_n^{\sbdm{\alp}}(z), \;\;\;\;\;
 g_n^B(\omg z) = \omg^{-1}g_n^B(z), \nonumber\\
 h_{{\rm R}n}^{(\pm)\sbdm{\mu}f}(\omg z) \eql \omg^{\mp\frac{1}{2}}e^{i\vph_f}
 e^{ip\cdot\sbdm{\mu}}h_{{\rm R}n}^{(\pm)\sbdm{\mu}f}(z), \;\;\;\;\;
 h_{{\rm L}n}^{(\pm)\sbdm{\mu}f}(\omg z) = \omg^{\pm\frac{1}{2}}e^{i\vph_f}
 e^{ip\cdot\sbdm{\mu}}h_{{\rm L}n}^{(\pm)\sbdm{\mu}f}(z). 
 \label{BC:md_eq:2}
\eea

The SS phases~$\phi_s=\phi_s^{\sbdm{\alp}},\phi_s^{\sbdm{\mu}f}$ 
in (\ref{BC:md_eq:1}) are defined modulo 1. 
This means that a set of solutions to the mode equation is invariant 
under $\phi_s\to\phi_s+1$. 
When $\abs{K}>1$ ($K=k_{\sbdm{\alp}},k_{\sbdm{\mu}f}$), however, 
each mode function is not invariant under such shifts. 
In fact, the shift:~$\phi_1\to\phi_1+1$ changes an eigenstate to another degenerate eigenstate, 
and the shift:~$\phi_\tau\to\phi_\tau+1$ rotates the phase of the mode function. 
(See Sec.~2.2 of Ref.~\cite{Abe:2013bca}.)
If we focus on a specific eigenstate among the degenerate mass eigenstates, 
the period of $\phi_s$ is $\abs{K}$, rather than 1. 

We should also note that 
the SS phases can be converted  
into the Wilson-line phases by a large gauge transformation, 
and vice versa~\cite{Abe:2013bca}.  
The correspondence is 
\bea
 \phi_s^{\sbdm{\alp}} = 0, \;\;\; \zt_{\sbdm{\alp}} & \exch & 
 \phi_s^{\sbdm{\alp}} = \frac{k_{\sbdm{\alp}}}{2\Im\tau}\Im(\bar{s}\zeta_{\sbdm{\alp}}), 
 \;\;\; \zeta_{\sbdm{\alp}} = 0, \nonumber\\
 \phi_s^{\sbdm{\mu}f} = 0, \;\;\; \zt_{\sbdm{\mu}f} & \exch & 
 \phi_s^{\sbdm{\mu}f} = \frac{k_{\sbdm{\mu}f}}{2\Im\tau}\Im(\bar{s}\zt_{\sbdm{\mu}f}), 
 \;\;\; \zt_{\sbdm{\mu}f} = 0,  \label{rel:WltoSS}
\eea
or equivalently, 
\bea
 \phi_s^{\sbdm{\alp}}, \;\;\;\zt_{\sbdm{\alp}} = 0 & \exch & 
 \phi_s^{\sbdm{\alp}} = 0, \;\;\;
 \zt_{\sbdm{\alp}} = \frac{2}{k_{\sbdm{\alp}}}
 \brkt{\tau\phi_1^{\sbdm{\alp}}-\phi_\tau^{\sbdm{\alp}}}, \nonumber\\
 \phi_s^{\sbdm{\mu}f}, \;\;\;\zt_{\sbdm{\mu}f} = 0 & \exch & 
 \phi_s^{\sbdm{\mu}f} = 0, \;\;\;
 \zt_{\sbdm{\mu}f} = \frac{2}{k_{\sbdm{\mu}f}}
 \brkt{\tau\phi_1^{\sbdm{\mu}f}-\phi_\tau^{\sbdm{\mu}f}}.  \label{rel:SStoWl}
\eea
In the following, we choose a gauge where all the SS phases are zero. 
As mentioned in Refs.~\cite{Abe:2013bca,Ibanez:1986tp,Kobayashi:1990mi,Abe:2009uz}, 
the Wilson-line phases can only take finite numbers 
(which are equal to the numbers of the orbifold fixed points) of values 
when the theory is compactified on $T^2/Z_N$ 
(see Appendix~\ref{C:anl_fm}).

\subsection{Mode equations}
We choose the following gauge-fixing terms:
\be
 \cL_{\rm gf} = -\frac{1}{2g_A^2}\tr\brc{\brkt{D^M\tl{A}_M}^2}
 -\frac{1}{2g_B^2}\brkt{\der^M\tl{B}_M}^2, 
\ee
where $\tl{A}_M\equiv A_M-\vev{A_M}$, $\tl{B}\equiv B_M-\vev{B_M}$, and 
\be
 D_M\tl{A}_N \equiv \der_M\tl{A}_N-i\sbk{\vev{A_M},\tl{A}_N}. 
\ee
Then, the mode equations are read off as 
\bea
 \der_z\der_{\bar{z}}f_n^i \eql -\tl{m}_n^2f_n^i, \;\;\;\;\;
 \cO_{\sbdm{\alp}}f_n^{\sbdm{\alp}} = -\tl{m}_n^2f_n^{\sbdm{\alp}}, \;\;\;\;\;
 \der_z\der_{\bar{z}}f_n^B = -\tl{m}_n^2f_n^B, \nonumber\\
 \der_z\der_{\bar{z}}g_n^i \eql -\tl{m}_n^2g_n^i, \;\;\;\;\;
 \brkt{\cO_{\sbdm{\alp}}+\frac{k_{\sbdm{\alp}}\pi}{2\Im\tau}}g_n^{\sbdm{\alp}} 
 = -\tl{m}_n^2g_n^{\sbdm{\alp}}, \;\;\;\;\;
 \der_z\der_{\bar{z}}g_n^B = -\tl{m}_n^2g_n^B, \nonumber\\
 D_{\bar{z}}^{(\sbdm{\mu}f)}h_{{\rm R}n}^{(+)\sbdm{\mu}f} 
 \eql -\tl{m}_nh_{{\rm L}n}^{(+)\sbdm{\mu}f}, \;\;\;\;\;
 D_z^{(\sbdm{\mu}f)}h_{{\rm L}n}^{(+)\sbdm{\mu}f}
 = \tl{m}_n^*h_{{\rm R}n}^{(+)\sbdm{\mu}f}, \nonumber\\
 D_z^{(\sbdm{\mu}f)}h_{{\rm R}n}^{(-)\sbdm{\mu}f}
 \eql -\tl{m}_nh_{{\rm L}n}^{(-)\sbdm{\mu}f}, \;\;\;\;\;
 D_{\bar{z}}^{(\sbdm{\mu}f)}h_{{\rm L}n}^{(-)\sbdm{\mu}f} 
 = \tl{m}_n^*h_{{\rm R}n}^{(-)\sbdm{\mu}f}, 
 \label{md_eq}
\eea
where $\tl{m}_n\equiv \pi R_1m_n$ ($m_n$ is the KK mass eigenvalues),\footnote{
The eigenvalues~$m_n$ are in general complex for the fermionic fields, 
while they are real for the bosonic fields because of the hermiticity 
of the corresponding differential operators. 
} 
and 
\bea
 \cO_{\sbdm{\alp}} \defa 
 \brkt{\der_{\bar{z}}+\frac{k_{\sbdm{\alp}}\pi(z+\zeta_{\sbdm{\alp}})}{2\Im\tau}}
 \brkt{\der_z-\frac{k_{\sbdm{\alp}}\pi(\bar{z}
 +\bar{\zeta}_{\sbdm{\alp}})}{2\Im\tau}}
 +\frac{k_{\sbdm{\alp}}\pi}{2\Im\tau} \nonumber\\
 \eql \brkt{\der_z-\frac{k_{\sbdm{\alp}}\pi(\bar{z}+\bar{\zeta}_{\sbdm{\alp}})}{2\Im\tau}}
 \brkt{\der_{\bar{z}}+\frac{k_{\sbdm{\alp}}\pi(z+\zeta_{\sbdm{\alp}})}{2\Im\tau}}
 -\frac{k_{\sbdm{\alp}}\pi}{2\Im\tau}, \nonumber\\
%
%
 D_z^{(\sbdm{\mu}f)} \defa \der_z-\frac{k_{\sbdm{\mu}f}\pi
 (\bar{z}+\bar{\zeta}_{\sbdm{\mu}f})}{2\Im\tau}, 
 \;\;\;\;\;
 D_{\bar{z}}^{(\sbdm{\mu}f)} \equiv \der_{\bar{z}}
 +\frac{k_{\sbdm{\mu}f}\pi(z+\zeta_{\sbdm{\mu}f})}{2\Im\tau}. 
\eea

\subsection{Mode functions on $\bdm{T^2}$}
Let us first find the mode functions defined on $T^2$, 
which are denoted by letters with tilde. 
They are obtained by solving (\ref{md_eq}) with (\ref{BC:md_eq:1}) 
in the manner of Refs.~\cite{Cremades:2004wa,Abe:2013bca}. 

\subsubsection{Gauge fields}
Since $C^i_M$ and $B_M$ do not feel the background gauge fields, 
their mode equations are easily solved, and the solutions are~\footnote{
For these modes, we label the KK level by a pair of integers.
} 
\bea
 \tl{f}_{n,l}^i(z), \; \tl{g}_{n,l}^i(z) \eql 
 \cN_{n,l}^{{\rm c}i}\cos\frac{2\pi\Im\brc{(n+l\bar{\tau})z}}{\Im\tau}
 +\cN_{n,l}^{{\rm s}i}\sin\frac{2\pi\Im\brc{(n+l\bar{\tau})z}}{\Im\tau}, \nonumber\\
 \tl{f}_{n,l}^B(z), \; \tl{g}_{n,l}^B(z) \eql 
 \cN_{n,l}^{{\rm c}B}\cos\frac{2\pi\Im\brc{(n+l\bar{\tau})z}}{\Im\tau}
 +\cN_{n,l}^{{\rm s}B}\sin\frac{2\pi\Im\brc{(n+l\bar{\tau})z}}{\Im\tau}, 
\eea
where $\cN_{n,l}^{{\rm c}i},\cN_{n,l}^{{\rm s}i},\cN_{n,l}^{{\rm c}B}$ 
and $\cN_{n,l}^{{\rm s}B}$ are real constants, 
and the corresponding mass eigenvalues are
\be
 \tl{m}_n = \frac{\pi\abs{n+l\tau}}{\Im\tau}. 
\ee
Note that the zero-mode functions are constant. 

For $W_M^{\sbdm{\alp}}$ with $k_{\sbdm{\alp}}=0$, 
the mode functions are affected only by the Wilson-line phases.\footnote{
Note that $k_{\sbdm{\alp}}\zeta_{\sbdm{\alp}}=Nc\cdot\bdm{\alp}/2\pi$ is 
independent of the flux~$\cC^i$. 
It can take nonvanishing values even in the case of $k_{\sbdm{\alp}}=0$. 
} 
\be
 \tl{f}_{n,l}^{\sbdm{\alp}}(z), \; \tl{g}_{n,l}^{\sbdm{\alp}}(z) = 
 \cN_{n,l}^{\sbdm{\alp}}\exp\brc{\frac{2\pi i}{\Im\tau}\Im\brc{\brkt{n+l\bar{\tau}
 -\frac{k_{\sbdm{\alp}}\bar{\zeta}_{\sbdm{\alp}}}{2}}z}}, 
\ee
where $\cN_{n,l}^{\sbdm{\alp}}$ are normalization constants, and 
the mass eigenvalues are
\be
 \tl{m}_{n,l} = \frac{\pi}{\Im\tau}\abs{n+l\tau-\frac{k_{\sbdm{\alp}}\zeta_{\sbdm{\alp}}}{2}}. 
 \label{KKmass:Wk0}
\ee

The other fields feel the magnetic fluxes,\footnote{
For simplicity, we do not consider the case of $k_{\sbdm{\mu}f}=0$. 
} 
and there are degenerate mass eigenstates at each KK level. 
For $W_\mu^{\sbdm{\alp}}$ with $k_{\sbdm{\alp}}\neq 0$, 
there are no zero-modes, \ie, 
\be
 \tl{m}_n^2 = \brkt{n+\frac{1}{2}}\frac{\abs{k_{\sbdm{\alp}}}\pi}{\Im\tau} 
 \geq \frac{\abs{k_{\sbdm{\alp}}}\pi}{2\Im\tau} > 0. 
\ee
As for $W_z^{\sbdm{\alp}}$, only components with $k_{\sbdm{\alp}}>0$ have zero-modes. 
The corresponding mode functions are 
\be
 \tl{g}_0^{\sbdm{\alp}(j)}(z) =  
 \cF^{(j)}(z;k_{\sbdm{\alp}},\zt_{\sbdm{\alp}}), 
\ee
where $j=1,2,\cdots,k_{\sbdm{\alp}}$, and 
\be
 \cF^{(j)}(z;K,\zt) \equiv \begin{cases} (2K\Im\tau)^{\frac{1}{4}}
 e^{K\pi i(z+\zt)\frac{\Im(z+\zt)}{\Im\tau}}
 \thF{\frac{j}{K}}{0}(K(z+\zt),K\tau), & (K>0) \\
 (2\abs{K}\Im\tau)^{\frac{1}{4}}
 e^{K\pi i(\bar{z}+\bar{\zt})\frac{\Im(\bar{z}+\bar{\zt})}{\Im\bar{\tau}}}
 \thF{\frac{j}{K}}{0}(K(\bar{z}+\bar{\zt}),K\bar{\tau}). & (K<0) \end{cases}
 \label{def:cF}
\ee
Here, $\thF{a}{b}$ is the Jacobi theta function defined by
\be
 \thF{a}{b}(Kz,K\tau) \equiv \sum_{l=-\infty}^\infty 
 e^{\pi i(l+a)^2K\tau}e^{2\pi i(l+a)(Kz+b)}. 
\ee
The function~$\cF^{(j)}$ satisfies the relation:
\be
 \brc{\cF^{(j)}(z;K,\zt)}^* = \cF^{(-j)}(z;-K,\zt), \label{cF:cc}
\ee
and is normalized as
\be
 \int_{T^2}\dr^2z\;\brc{\cF^{(j)}(z;K,\zeta)}^*\cF^{(k)}(z;K,\zeta) = \dlt_{jk}. 
 \label{cF:orthnml}
\ee
The mode functions for the KK excitation modes are 
\be
 \tl{g}_n^{\sbdm{\alp}(j)}(z) \propto \brkt{D_z^{(\sbdm{\alp})}}^n
 \tl{g}_0^{\sbdm{\alp}(j)}(z), 
\ee
where 
\be
 D_z^{(\sbdm{\alp})} \equiv \der_z
 -\frac{k_{\sbdm{\alp}}\pi(\bar{z}+\bar{\zt}_{\sbdm{\alp}})}{2\Im\tau}. 
\ee
and the mass eigenvalues are 
\be
 \tl{m}_n^2 = \frac{nk_{\sbdm{\alp}}\pi}{\Im\tau}. 
\ee

The components of $W_z^{\sbdm{\alp}}$ with $k_{\sbdm{\alp}}<0$ do not have zero-modes, and 
\be
 \tl{m}_n^2 = \frac{(n+1)\abs{k_{\sbdm{\alp}}}\pi}{\Im\tau} 
 \geq \frac{\abs{k_{\sbdm{\alp}}}\pi}{\Im\tau} > 0. 
\ee

\subsubsection{Fermions}
For components of $\Psi^f_{\chi_6}$ with $k_{\sbdm{\mu}f}>0$, 
only $\psi_+^f$ and $\bar{\lmd}_-^f$ have zero-modes whose mode functions are given by 
\be
 \tl{h}_{{\rm R}0}^{(+)\sbdm{\mu}f(j)}(z), \;
 \tl{h}_{{\rm L}0}^{(-)\sbdm{\mu}f(j)}(z) = 
 \cF^{(j)}(z;k_{\sbdm{\mu}f},\zt_{\sbdm{\mu}f}), 
\ee
where $j=1,2,\cdots,k_{\sbdm{\mu}f}$. 
For components of $\Psi^f_{\chi_6}$ with $k_{\sbdm{\mu}f}<0$, 
only $\psi_-^f$ and $\bar{\lmd}_+^f$ have zero-modes whose mode functions are
\be
 \tl{h}_{{\rm R}0}^{(-)\sbdm{\mu}f(j)}(z), \;
 \tl{h}_{{\rm L}0}^{(+)\sbdm{\mu}f(j)}(z) = 
 \cF^{(j)}(z;k_{\sbdm{\mu}f},\zt_{\sbdm{\mu}f}), 
\ee
where $j=1,2,\cdots,\abs{k_{\sbdm{\mu}f}}$. 

The mode functions for the KK excitation modes are obtained by operating 
$D_z^{(\sbdm{\mu}f)}$ (for $k_{\sbdm{\mu}f}>0$) or 
$D_{\bar{z}}^{(\sbdm{\mu}f)}$ (for $k_{\sbdm{\mu}f}<0$) 
on the above functions, and their mass eigenvalues are
\be
 \tl{m}_n^2 = \frac{n\abs{k_{\sbdm{\mu}f}}\pi}{\Im\tau}. 
\ee

\subsection{Mode functions on $\bdm{T^2/Z_N}$}
As we have seen in the previous subsection, $\{\tl{f}_0^i(z),\tl{f}_0^B(z)\}$ 
and $\{\tl{g}_0^i(z),\tl{g}_0^B(z)\}$ are constants. 
The former satisfies the orbifold boundary conditions in (\ref{BC:md_eq:2}), 
but the latter does not. 
Thus, $C_\mu^i$ and $B_\mu$ have zero-modes on $T^2/Z_N$ 
while $C_z^i$ and $B_z$ do not. 

As for $W_M^{\sbdm{\alp}}$ with $k_{\sbdm{\alp}}=0$, 
zero-modes exist on $T^2$ only when $\zeta_{\sbdm{\alp}}=0$ 
(see (\ref{KKmass:Wk0})).\footnote{
Note that $k_{\sbdm{\alp}}\zt_{\sbdm{\alp}}/2$ is defined 
modulo 1 and $\tau$ as can be seen from (\ref{rel:SStoWl}). 
} 
Since the corresponding mode functions are constants, 
they satisfy (\ref{BC:md_eq:2}) only when $p\cdot\bdm{\alp}=0$ 
for $f_0^{\sbdm{\alp}}(z)$, 
and $p\cdot\bdm{\alp}=2\pi/N$ for $g_0^{\sbdm{\alp}}(z)$. 
These are the conditions for $W_\mu^{\sbdm{\alp}}$ and $W_z^{\sbdm{\alp}}$ 
have zero-modes on $T^2/Z_N$. 

The other modes feel the magnetic fluxes. 
Thus, they have degenerate modes at each KK level. 
The orbifold boundary conditions in (\ref{BC:md_eq:2}) have the form
\be
 F_n^{(j)}(\omg z) = \eta F_n^{(j)}(z), \label{F:orbifoldBC}
\ee
where $\eta$ is an $N$-th root of unity, 
and $j=1,2,\cdots,\abs{K}$ discriminates the degenerate modes. 
Note that
\be
 \hat{F}_0^{(j)}(z) \equiv \frac{1}{N}\sum_{l=0}^{N-1}\eta^{-l}\tl{F}_0^{(j)}(\omg^l z), 
 \label{def:hatF}
\ee
where $\tl{F}_0^{(j)}(z)$ is a zero-mode function on $T^2$, 
satisfies (\ref{F:orbifoldBC}). 
Since $\tl{F}_0^{(j)}(\omg^lz)$ is a solution of 
(\ref{md_eq}) that satisfies (\ref{BC:md_eq:1}), it can be expressed 
as a linear combination of $\tl{F}_0^{(j)}(z)$, \ie, 
\be
 \tl{F}_0^{(j)}(\omg^lz) = \sum_{k=1}^{\abs{K}}D_{jk}^{(\omg^l)}
 \tl{F}_0^{(k)}(z), 
 \label{def:C_jk} 
\ee
where $D_{jk}^{(\omg^l)}$ are constants. 
Thus, $\hat{F}_0^{(j)}(z)$ in (\ref{def:hatF}) is expressed as
\be
 \hat{F}_0^{(j)}(z) = \sum_{k=1}^{\abs{K}}\cM_{jk}^{(\eta)}\tl{F}_0^{(k)}(z), 
 \label{expr:hatF}
\ee
where 
\be
 \cM_{jk}^{(\eta)} \equiv \frac{1}{N}\sum_{l=0}^{N-1}\eta^{-l}D_{jk}^{(\omg^l)}. 
 \label{def:D^eta}
\ee
Although $j$ runs from 1 to $\abs{K}$, 
not all of $\hat{F}_0^{(j)}(z)$ are independent mode functions~\cite{Abe:2013bca}. 
In fact, the matrix~$\cM^{(\eta)}$ generically has zero eigenvalues. 
The number of zero-modes is equal to the rank of $\cM^{(\eta)}$. 
Here, note that the matrix~$\cM^{(\eta)}$ is hermitian because 
\be
 \cM^{(\eta)\dagger} = \frac{1}{N}\sum_{l=0}^{N-1}\eta^lD^{(\bar{\omg}^l)\dagger} 
 = \frac{1}{N}\sum_{l'=0}^{N-1}\eta^{-l'}D^{(\omg^{l'})} 
 = \cM^{(\eta)}, 
\ee
where $l'\equiv -l$. 
(See Appendix~\ref{C:anl_fm}.)
Thus, $\cM^{(\eta)}$ can be diagonalized by a unitary matrix~$V^{(\eta)}$: 
\be
 V^{(\eta)}\cM^{(\eta)}V^{(\eta)\dagger} = \diag\brkt{\lmd_1,\lmd_2,\cdots,\lmd_r,0,\cdots,0}, 
\ee
where $\lmd_j$ ($j=1,2,\cdots,r$) are the non-zero (real) eigenvalues, 
and $r\equiv {\rm Rank}\,\cM^{(\eta)}$. 
Then we find that
\be
 \sum_{k=1}^{\abs{K}} V_{jk}^{(\eta)}\hat{F}_0^{(k)}(z) = \begin{cases} 
 \lmd_j\sum_k V_{jk}^{(\eta)}\tl{F}_0^{(k)}(z), & (1\leq j\leq r) \\
 0. & (r+1\leq j\leq \abs{K}) \end{cases}
\ee
Therefore, it is convenient to choose independent mode functions on $T^2/Z_N$ as
\be
 F_0^{(j)}(z) \equiv \sqrt{N}\sum_{k=1}^{\abs{K}}V_{jk}^{(\eta)}
 \tl{F}_0^{(k)}(z), \label{def:Forbifold}
\ee
where $j=1,2,\cdots,r$. 
We can easily show that these satisfy the orthonormal condition:
\be
 \int_{T^2/Z_N}\dr^2z\;\brc{F_0^{(j)}(z)}^*F_0^{(k)}(z) = \dlt_{jk}, 
 \label{F:othnml}
\ee
which follows from the orthonormal condition of $\tl{F}_n^{(j)}(z)$. 
The matrix~$\cM^{(\eta)}$ is expressed as
\be
 \cM_{jk}^{(\eta)} = \int_{T^2}\dr^2z\;\brc{\tl{F}_0^{(j)}(z)}^*\hat{F}_0^{(k)}(z). 
\ee
In Ref.~\cite{Abe:2014noa}, analytic forms of the matrix~$\cM^{(\eta)}$ are derived 
implying the operator formalism. 
It is obtained from (\ref{def:D^eta}) with analytic forms of $D_{jk}^{(\omg^l)}$, 
which are collected in Appendix~\ref{C:anl_fm}. 

The mode functions for the KK modes are obtained by operating 
$D_z=D_z^{(\sbdm{\alp})},D_z^{(\sbdm{\mu}f)}$ 
or $D_{\bar{z}}=D_{\bar{z}}^{(\sbdm{\alp})},D_{\bar{z}}^{(\sbdm{\mu}f)}$ 
on $F_0^{(j)}(z)$, just like those on $T^2$. 
However, since 
\bea
 D_z\brkt{\tl{F}_0^{(j)}(\omg^l z)} \eql 
 \omg^l \brkt{D_z\tl{F}_0^{(j)}}(\omg^l z)
 \propto \omg^l\tl{F}_1^{(j)}(\omg^lz), \;\;\;\;\;
 (\mbox{when $K>0$}) \nonumber\\
 D_{\bar{z}}\brkt{\tl{F}_0^{(j)}(\omg^l z)} \eql 
 \bar{\omg}^l\brkt{D_{\bar{z}}\tl{F}_0^{(j)}}(\omg^l z) 
 \propto \bar{\omg}^l\tl{F}_1^{(j)}(\omg^l z), \;\;\;\;\;
 (\mbox{when $K<0$})
\eea
the phase factor~$\eta$ in $\cM_{jk}^{(\eta)}$ becomes 
$\eta\omg^{-1}$ (for $K>0$) or $\eta\omg$ (for $K<0$). 
Therefore, the expression corresponding to (\ref{def:Forbifold}) 
for the KK modes is
\be
 F_n^{(j)}(z) = \begin{cases} \displaystyle
 \sqrt{N}\sum_{k=1}^{K}V_{jk}^{(\eta\omg^{-n})}
 \tl{F}_n^{(k)}(z) & (\mbox{for $K>0$}) \\
 \displaystyle
 \sqrt{N}\sum_{k=1}^{\abs{K}}V_{jk}^{(\eta\omg^n)}
 \tl{F}_n^{(k)}(z) & (\mbox{for $K<0$}) \end{cases}. 
\ee
The number of mass eigenstates at each KK level is given by 
the rank of $\cM^{(\eta\omg^{-n})}$ (for $K>0$) or 
that of $\cM^{(\eta\omg^n)}$ (for $K<0$). 

Note that the constants~$D^{(\omg^l)}_{jk}$ in Appendix~\ref{C:anl_fm}, 
which are functions of $K$ and $\zt$, satisfy 
\be
 D^{(\omg^l)}_{jk}[-K,\zt] = D^{(\bar{\omg}^l)}_{kj}[K,\zt], 
\ee
where $\zt=\frac{2}{K}(\tau\phi_1-\phi_\tau)$. 
Thus, we find that
\bea
 \cM^{(\eta)}_{jk}[-K,\zt] \eql 
 \frac{1}{N}\sum_{l=0}^{N-1}\bar{\eta}^lD^{(\omg^l)}_{jk}[-K,\zt] 
 = \frac{1}{N}\sum_{l=0}^{N-1}\bar{\eta}^l D^{(\bar{\omg}^l)}_{kj}[K,\zt] \nonumber\\
 \eql \frac{1}{N}\sum_{l'=0}^{N-1}\bar{\eta}^{-l'}
 D^{(\omg^{l'})}_{kj}[K,\zt] = \cM^{(\bar{\eta})}_{kj}[K,\zt], 
\eea
where $l'\equiv -l$. 
This indicates that the number of zero-modes for a field that feels 
a magnetic flux~$K<0$ and the orbifold twist phase~$\eta$ is equal to 
that for a field with $\abs{K}$ and $\bar{\eta}$.

\section{Yukawa coupling constants} \label{Yukawa_cp}
\subsection{General expression}
In the gauge-Higgs unification, the Yukawa couplings originate 
from the 6D gauge interactions:
\bea
 S \eql \int\dr^6x\;\brkt{\sum_{f_+}i\bar{\Psi}_+^{f_+}\gm^M\cD_M\Psi_+^{f_+}
 +\sum_{f_-}i\bar{\Psi}_-^{f_-}\gm^M\cD_M\Psi_-^{f_-}}+\cdots 
 \nonumber\\
 \eql \int\dr^4x\int\dr^2z\;2\pi R_1\brkt{-\sum_{f_+}
 i\bar{\psi}_+^{f_+}A_z\bar{\lmd}_+^{f_+}
 +\sum_{f_-}i\lmd_-^{f_-}A_z\psi_-^{f_-}}+\hc+\cdots,  
\eea
where $d^2z\equiv dzd\bar{z}$. 
In the 4D effective theory, we have the following Yukawa couplings:
\bea
 \cL_{\rm yukawa}^{(4D)} \eql \sum_{\sbdm{\mu}}\sum_{f_+}\sum_{i,j,k}
 y_{ijk}^{(+)\sbdm{\mu}f_+}
 \bar{\psi}_{+0}^{(\sbdm{\mu}+\sbdm{\alp})f_+(i)}
 \vph_0^{\sbdm{\alp}(k)}\bar{\lmd}_{+0}^{\sbdm{\mu}f_+(j)}+\hc \nonumber\\
 &&+\sum_{\sbdm{\mu}}\sum_{f_-}\sum_{i,j,k}
 y^{(-)\sbdm{\mu}f_-}_{ijk}
 \lmd_{-0}^{\sbdm{\mu}f_-(j)}\vph_0^{\sbdm{\alp}(k)}
 \psi_{-0}^{(\sbdm{\mu}+\sbdm{\alp})f_-(i)}+\hc, 
\eea
where the indices~$i,j,k$ run over the degenerate zero-modes, and 
\bea
 y_{ijk}^{(+)\sbdm{\mu}f_+} \defa -\frac{ig_A
 \langle\bdm{\mu}+\bdm{\alp}|E_{\sbdm{\alp}}|\bdm{\mu}\rangle}{\pi R_1}
 \int_{T^2/Z_N}\dr^2z\;\brc{h_{{\rm R}0}^{(+)(\sbdm{\mu}+\sbdm{\alp})f_+(i)}(z)}^* 
 g_0^{\sbdm{\alp}(k)}(z)h_{{\rm L}0}^{(+)\sbdm{\mu}f_+(j)}(z) \nonumber\\
 \eql -\frac{2i\bar{g}_A\sqrt{\Im\tau}}{N\sqrt{N}}
 \langle\bdm{\mu}+\bdm{\alp}|E_{\sbdm{\alp}}|\bdm{\mu}\rangle
 \int_{T^2}\dr^2z\;
 \brc{h_{{\rm R}0}^{(+)(\sbdm{\mu}+\sbdm{\alp})f_+(i)}(z)}^* 
 g_0^{\sbdm{\alp}(k)}(z)h_{{\rm L}0}^{(+)\sbdm{\mu}f_+(j)}(z) \nonumber\\
 \eql -2i\bar{g}_A\sqrt{\Im\tau}\langle\bdm{\mu}+\bdm{\alp}|E_{\sbdm{\alp}}|\bdm{\mu}\rangle
 \sum_{i'=1}^{\abs{K_1}}\sum_{j'=1}^{\abs{K_2}}
 \sum_{k'=1}^{\abs{K_3}}
 V_{ii'}^{(\eta_1)*}V_{jj'}^{(\eta_2)}V_{kk'}^{(\eta_3)} \nonumber\\
 &&\times\int_{T^2}\dr^2z\;
 \cF^{(i')*}(z;K_1,\zt_1)\cF^{(j')}(z;K_2,\zt_2)\cF^{(k')}(z;K_3,\zt_3), 
 \nonumber\\
 y_{ijk}^{(-)\sbdm{\mu}f_-} \defa 2i\bar{g}_A\sqrt{\Im\tau}
 \langle\bdm{\mu}+\bdm{\alp}|E_{\sbdm{\alp}}|\bdm{\mu}\rangle
 \sum_{i'=1}^{\abs{K_1}}\sum_{j'=1}^{\abs{K_2}}\sum_{k'=1}^{\abs{K_3}}
 V_{ii'}^{(\eta_1)}V_{jj'}^{(\eta_2)*}V_{kk'}^{(\eta_3)} \nonumber\\
 &&\times\int_{T^2}\dr^2z\;
 \cF^{(i')}(z;K_1,\zt_1)\cF^{(j')*}(z;K_2,\zt_2)\cF^{(k')}(z;K_3,\zt_3), 
\eea
where $\bar{g}_A\equiv\frac{g_A}{\sqrt{\cA}}=\frac{\sqrt{N}g_A}{2\pi R_1\sqrt{\Im\tau}}$ 
is the 4D gauge coupling constant, 
$K_1\equiv k_{(\sbdm{\mu+\alp})f_\pm}$, $\zt_1\equiv \zt_{(\sbdm{\mu+\alp})f_\pm}$, 
$K_2\equiv k_{\sbdm{\mu}f_\pm}$, $\zt_2\equiv \zt_{\sbdm{\mu}f_\pm}$, 
$K_3\equiv k_{\sbdm{\alp}}$, $\zt_3\equiv \zt_{\sbdm{\alp}}$, 
and $\{\eta_1,\eta_2,\eta_3\}$ are the phase factors in the orbifold 
boundary conditions.\footnote{  
The phase factors~$\eta_1$ and $\eta_2$ depend on the flavor index~$f_+$ or $f_-$. 
}

\subsubsection{Couplings to fermions with $\bdm{\chi_6=+}$}
From the gauge invariance of the Lagrangian, the following conditions hold:
\be
 K_1 = K_2+K_3, \;\;\;\;\;
 K_1\zt_1 = K_2\zt_2+K_3\zt_3, \label{rel:gauge_inv}
\ee
and from the condition that the zero-modes exist, 
it follows that 
\be
 K_1>0, \;\; K_2 <0, \;\; K_3 > 0. \label{K:signs}
\ee
Then we find that
\bea
 &&\brc{\cF^{(i')}(z;K_1,\zt_1)}^*\cF^{(j')}(z;K_2,\zt_2) \nonumber\\
 \eql \frac{1}{\sqrt{K_3}}\sum_{m=1}^{K_3}
 \brc{\cF^{(i'-j'+K_1m)}\brkt{z;K_3,\zt_3}
 \cF^{(\abs{K_2}i'+K_1j'+K_1\abs{K_2}m)}
 \brkt{0;\abs{K_1K_2K_3},\frac{\zt_1-\zt_2}{K_3}}}^*, \nonumber\\
\eea
which follows from the formula ((5.8) in Ref.~\cite{Cremades:2004wa}):
\bea
 &&\thF{\frac{i'}{K_1}}{0}\brkt{K_1(z+\zt_1),K_1\tau}\cdot
 \thF{-\frac{j'}{\abs{K_2}}}{0}\brkt{\abs{K_2}(z+\zt_2),\abs{K_2}\tau} \nonumber\\
 \eql \sum_{l=1}^{K_1+\abs{K_2}}\thF{\frac{i'-j'+K_1l}{K_1+\abs{K_2}}}{0}
 \brkt{(K_1+\abs{K_2})\brkt{z+\frac{K_1\zt_1+\abs{K_2}\zt_2}{K_1+\abs{K_2}}},
 (K_1+\abs{K_2})\tau} \nonumber\\
 &&\times\thF{\frac{\abs{K_2}i'+K_1j'+K_1\abs{K_2}l}{K_1\abs{K_2}(K_1+\abs{K_2})}}{0}
 \brkt{K_1\abs{K_2}(\zt_1-\zt_2),K_1\abs{K_2}(K_1+\abs{K_2})\tau}, 
\eea
with (\ref{cF:cc}) and (\ref{rel:gauge_inv}). 
Therefore, using the orthonormal condition~(\ref{cF:orthnml}), we obtain 
\bea
 &&\int_{T^2}\dr^2z\;\brc{\cF^{(i')}(z;K_1,\zt_1)}^*
 \cF^{(j')}(z;K_2,\zt_2)\cF^{(k')}(z;K_3,\zt_3) \nonumber\\
 \eql \frac{1}{\sqrt{K_3}}\sum_{m=1}^{K_3}
 \cF^{(K_2i'-K_1j'+K_1K_2m)}\brkt{0,K_1K_2K_3,\frac{\zt_1-\zt_2}{K_3}}
 \dlt_{i'-j'+K_1m,k'}. 
\eea
Notice that $\dlt_{i'-j'+K_1m,k'}$ is defined on $Z_{K_3}$, \ie, 
\be
 \dlt_{i'-j'+K_1m,k'} \equiv \begin{cases} 1 & (i'-j'+K_1m=k' \mbox{ mod $K_3$}) \\
 0 & (\mbox{other cases}) \end{cases}.
\ee

As a result, we obtain the following expression for the Yukawa coupling constant: 
\bea
 y_{ijk}^{(+)} \eql -\frac{2i\bar{g}_A\sqrt{\Im\tau}}{\sqrt{K_3}}
 \langle\bdm{\mu}+\bdm{\alp}|E_{\sbdm{\alp}}|\bdm{\mu}\rangle
 \sum_{i'=1}^{K_1}\sum_{j'=1}^{\abs{K_2}}\sum_{k'=1}^{K_3}
 V_{ii'}^{(\eta_1)*}[K_1,\zt_1]V_{jj'}^{(\eta_2)}[K_2,\zt_2]V_{kk'}^{(\eta_3)}[K_3,\zt_3] 
 \nonumber\\
 &&\times\sum_{m=1}^{K_3}
 \cF^{(K_2i'-K_1j'+K_1K_2m)}\brkt{0,K_1K_2K_3,\frac{\zt_1-\zt_2}{K_3}}
 \dlt_{i'-j'+K_1m,k'}.  \label{expr:y_ijk}
\eea
Note that the matrix~$V^{(\eta)}$ depends on the flux and the Wilson-line phase. 
The indices~$i$, $j$ and $k$ run from 1 to ${\rm Rank}\,\cM^{(\eta_1)}$, 
${\rm Rank}\,\cM^{(\eta_2)}$ and ${\rm Rank}\,\cM^{(\eta_3)}$, respectively.

\subsubsection{Couplings to fermions with $\bdm{\chi_6=-}$}
From the gauge invariance, $K_a$ and $\zt_a$ ($a=1,2,3$) satisfy 
\be
 K_2 = K_1+K_3, \;\;\;\;\;
 K_2\zt_2 = K_1\zt_1+K_3\zt_3, 
\ee
and the zero-mode conditions are 
\be
 K_1 < 0, \;\; K_2 > 0, \;\; K_3 > 0. 
\ee
Following the same procedure in the previous case, we obtain 
\bea
 y_{ijk}^{(-)} \eql \frac{2i\bar{g}_A\sqrt{\Im\tau}}{\sqrt{K_3}}
 \langle\bdm{\mu}+\bdm{\alp}|E_{\sbdm{\alp}}|\bdm{\mu}\rangle
 \sum_{i'=1}^{\abs{K_1}}\sum_{j'=1}^{K_2}\sum_{k'=1}^{K_3}
 V_{ii'}^{(\eta_1)}[K_1,\zt_1]V_{jj'}^{(\eta_2)*}[K_2,\zt_2]V_{kk'}^{(\eta_3)}[K_3,\zt_3] 
 \nonumber\\
 &&\times\sum_{m=1}^{K_3}
 \cF^{(K_1 j'-K_2 i'+K_1K_2m)}\brkt{0,K_1K_2K_3,\frac{\zt_2-\zt_1}{K_3}}
 \dlt_{j'-i'+K_2m,k'}.  \label{expr:y_ijk:2}
\eea

\subsection{Specific model}
In this subsection, we evaluate the Yukawa coupling constants 
in a specific model. 
We consider the case that $G={\rm SU(3)}$,\footnote{
We do not consider the custodial symmetry, for simplicity.  
}
$N=3$, 
and the matter fermions consist of 
two $\chi_6=-$ spinors~$(\Psi_-^1,\Psi_-^3)$ 
that belong to $\bdm{3}$ of ${\rm SU(3)}$
and two $\chi_6=+$ spinors~$(\Psi_+^2,\Psi_+^4)$ that belong to $\bdm{\bar{3}}$. 
The ${\rm U(1)}_X$ charges are assigned as 
$(q_1,q_2,q_3,q_4)=(0,1/3,-2/3,-1/3)$.

\subsubsection{Symmetry breaking and irreducible decomposition}
The roots of ${\rm SU(3)}$ are 
\bea
 &&\bdm{\alp}_1 \equiv \brkt{\frac{1}{2},\frac{\sqrt{3}}{2}}, \;\;\;
 \bdm{\alp}_2 \equiv \brkt{\frac{1}{2},-\frac{\sqrt{3}}{2}}, \;\;\;
 \bdm{\alp}_3 \equiv \bdm{\alp}_1+\bdm{\alp}_2 = (1,0), \nonumber\\
 &&-\bdm{\alp}_1, \;\;\; -\bdm{\alp}_2, \;\;\; -\bdm{\alp}_3, 
 \label{roots}
\eea
the weights of $\bdm{3}$ are 
\bea
 &&\bdm{\mu}_1 \equiv \brkt{\frac{1}{2},\frac{1}{2\sqrt{3}}}, \;\;\;
 \bdm{\mu}_2 \equiv \bdm{\mu}_1-\bdm{\alp}_1 = \brkt{0,-\frac{1}{\sqrt{3}}}, \nonumber\\
 &&\bdm{\mu}_3 \equiv \bdm{\mu}_1-\bdm{\alp}_1-\bdm{\alp}_2 
 = \brkt{-\frac{1}{2},\frac{1}{2\sqrt{3}}}, 
 \label{weights}
\eea
and the weights of $\bdm{\bar{3}}$ are 
$\{-\bdm{\mu}_1,-\bdm{\mu}_2,-\bdm{\mu}_3\}$. 

We choose the direction of the $G$ flux in (\ref{bg:AB_z}) as 
\be
 (\cC^1,\cC^2) = \cC^1\brkt{1,-\frac{1}{\sqrt{3}}}, 
\ee
so that $G$ is broken to ${\rm SU(2)}_{\rm L}\times {\rm U(1)}_Z$. 
Then, $\bdm{\alp}_{\rm L}$ and $\bdm{\eta}$ in (\ref{def:alp_L}) are identified as
\be
 \bdm{\alp}_{\rm L} = \bdm{\alp}_1, \;\;\;\;\;
 \bdm{\eta} = \brkt{\frac{1}{2},-\frac{1}{2\sqrt{3}}}. 
\ee
The normalization of $\bdm{\eta}$ is chosen in such a manner that the hypercharge of 
the Higgs doublet becomes $\pm 1/2$ (see (\ref{adjoint_decomp})). 
The fluxes~$\cC^1$ and $\cB$ are determined so that the quantization 
condition~(\ref{quantize:CB}) is satisfied for all the roots and the weights. 
In this model, (\ref{quantize:CB}) becomes 
\bea
 0 \eql 2k_{\pm\sbdm{\alp}_1}\pi, \;\;\;\;\;
 \pm N\cC^1 = 2k_{\pm\sbdm{\alp}_2}\pi = 2k_{\pm\sbdm{\alp}_3}\pi, \nonumber\\
 \frac{N\cC^1}{3} \eql 2k_{\sbdm{\mu}_11}\pi = 2k_{\sbdm{\mu}_21}\pi, \;\;\;\;\;
 -\frac{2N\cC^1}{3} = 2k_{\sbdm{\mu}_31}\pi, \nonumber\\
 N\brkt{-\frac{\cC^1}{3}+\frac{\cB}{3}} \eql 2k_{-\sbdm{\mu}_12}\pi 
 = 2k_{-\sbdm{\mu}_22}\pi, \;\;\;\;\;
 N\brkt{\frac{2\cC^1}{3}+\frac{\cB}{3}} = 2k_{-\sbdm{\mu}_32}\pi, \nonumber\\
 N\brkt{\frac{\cC^1}{3}-\frac{2\cB}{3}} \eql 2k_{\sbdm{\mu}_13}\pi 
 = 2k_{\sbdm{\mu}_23}\pi, \;\;\;\;\;
 N\brkt{-\frac{2\cC^1}{3}-\frac{2\cB}{3}} = 2k_{\sbdm{\mu}_33}\pi, \nonumber\\
 N\brkt{-\frac{\cC^1}{3}-\frac{\cB}{3}} \eql 2k_{-\sbdm{\mu}_14}\pi 
 = 2k_{-\sbdm{\mu}_24}\pi, \;\;\;\;\;
 N\brkt{\frac{2\cC^1}{3}-\frac{\cB}{3}} = 2k_{-\sbdm{\mu}_34}\pi. 
\eea
These can be solved as
\bea
 N\cC^1 \eql 6\kp\pi, \;\;\;\;\;
 N\cB = 6\kp'\pi, \nonumber\\
 k_{\pm\sbdm{\alp}_1} \eql 0, \;\;\;\;\;
 k_{\pm\sbdm{\alp}_2} = k_{\pm\sbdm{\alp}_3} = \pm 3\kp, \nonumber\\
 k_{\sbdm{\mu}_11} \eql k_{\sbdm{\mu}_21} = \kp, \;\;\;\;\;
 k_{\sbdm{\mu}_31} = -2\kp, \nonumber\\
 k_{-\sbdm{\mu}_12} \eql k_{-\sbdm{\mu}_22} = -\kp+\kp', \;\;\;\;\;
 k_{-\sbdm{\mu}_32} = 2\kp+\kp', \nonumber\\
 k_{\sbdm{\mu}_13} \eql k_{\sbdm{\mu}_23} = \kp-2\kp', \;\;\;\;\;
 k_{\sbdm{\mu}_33} = -2\kp-2\kp', \nonumber\\
 k_{-\sbdm{\mu}_14} \eql k_{-\sbdm{\mu}_24} = -\kp-\kp', \;\;\;\;\;
 k_{-\sbdm{\mu}_34} = 2\kp-\kp', 
 \label{sol:kkp}
\eea
where $\kp$ and $\kp'$ are integers. 

Under the unbroken ${\rm SU(2)}_{\rm L}$, the SU(3) adjoint representation is 
decomposed as 
\bea
 \brc{|-\bdm{\alp}_1\rangle, \; |\bdm{0}\rangle_T, \;|\bdm{\alp}_1\rangle} 
 &:& \mbox{triplet ($Y=0$)} \nonumber\\
 \brc{|\bdm{\alp}_2\rangle, \; |\bdm{\alp}_3\rangle} &:& 
 \mbox{doublet ($Y=1/2$)} \nonumber\\
 \brc{|-\bdm{\alp}_3\rangle, \; |-\bdm{\alp}_2\rangle} &:& 
 \mbox{doublet ($Y=-1/2$)} \nonumber\\
 |\bdm{0}\rangle_S &:& \mbox{singlet ($Y=0$)} \label{adjoint_decomp}
\eea
where $|\bdm{0}\rangle_T$ and $|\bdm{0}\rangle_S$ are the states 
that correspond to the Cartan generators, and $Y$ is the hypercharge. 
Since the above states do not have the ${\rm U(1)}_X$ charges, $Y$ in (\ref{adjoint_decomp}) 
is equal to the ${\rm U(1)}_Z$ charge. 
Thus, the Higgs doublets are identified as 
$(\vph_0^{\sbdm{\alp}_2(k)},\vph_0^{\sbdm{\alp}_3(k)})$ or
$(\vph_0^{-\sbdm{\alp}_3(k)},\vph_0^{-\sbdm{\alp}_2(k)})$. 

As for the matter sector, $\brc{|\bdm{\mu}_2\rangle,|\bdm{\mu}_1\rangle}$ 
and $\brc{|-\bdm{\mu}_1\rangle,|-\bdm{\mu}_2\rangle}$  
($|\bdm{\mu}_3\rangle$ and $|-\bdm{\mu}_3\rangle$) are doublets (singlets) 
of ${\rm SU(2)}_{\rm L}$. 
From (\ref{def:Y}), the hypercharges of the components of $\Psi_-^{1,3}$ are 
\bea
 \brkt{Y(\bdm{\mu}_1),Y(\bdm{\mu}_2),Y(\bdm{\mu}_3)} \eql 
 \brkt{\bdm{\eta}\cdot\bdm{\mu}_1,\bdm{\eta}\cdot\bdm{\mu}_2,\bdm{\eta}\cdot\bdm{\mu}_3}
 +(q_f,q_f,q_f) \nonumber\\
 \eql \begin{cases} \displaystyle 
 \brkt{\frac{1}{6},\frac{1}{6},-\frac{1}{3}} & (\mbox{for $\Psi_-^1$}) \\[3mm] \displaystyle
 \brkt{-\frac{1}{2},-\frac{1}{2},-1} & (\mbox{for $\Psi_-^3$}) \end{cases}, 
\eea
and those of $\Psi_+^{2,4}$ are 
\be
 \brkt{Y(-\bdm{\mu}_1),Y(-\bdm{\mu}_2),Y(-\bdm{\mu}_3)} 
 = \begin{cases} \displaystyle 
 \brkt{\frac{1}{6},\frac{1}{6},\frac{2}{3}} & (\mbox{for $\Psi_+^2$}) \\[3mm] \displaystyle
 \brkt{-\frac{1}{2},-\frac{1}{2},0} & (\mbox{for $\Psi_+^4$}) \end{cases}. 
\ee
Thus $(\bar{\lmd}_{- 0}^{\sbdm{\mu}_2 f(j)},\bar{\lmd}_{- 0}^{\sbdm{\mu}_1 f(j)})$ 
and $(\bar{\lmd}_{+ 0}^{-\sbdm{\mu}_1 f(j)},\bar{\lmd}_{+ 0}^{-\sbdm{\mu}_2 f(j)})$ 
($\psi_{- 0}^{\sbdm{\mu}_3f(i)}$ and $\psi_{+ 0}^{-\sbdm{\mu}_3f(i)}$) 
are identified as the left-handed doublets (the right-handed singlets) 
in the standard model. 
They are denoted by 
\bea
 &&Q_L^j(\bdm{2_{1/6}}), \;\;\; d_R^i(\bdm{1_{-1/3}}), \;\;\; (\mbox{from $\Psi_-^1$}) 
 \nonumber\\
 &&Q^{\prime j}_L(\bdm{2_{1/6}}), \;\;\; u^i_R(\bdm{1_{2/3}}), \;\;\; (\mbox{from $\Psi_+^2$}) 
 \nonumber\\
 &&L^j_L(\bdm{2_{-1/2}}), \;\;\; e_R^i(\bdm{1_{-1}}), \;\;\; (\mbox{from $\Psi_-^3$}) 
 \nonumber\\
 &&L^{\prime j}_L(\bdm{2_{-1/2}}), \;\;\; \nu_R^i(\bdm{1_0}), \;\;\; (\mbox{from $\Psi_+^4$})
 \label{QLcomp}
\eea
where $L$ and $R$ denote the 4D chiralities.

\subsubsection{Model parameters} \label{model_prmt}
We choose the matrix~$P$ in (\ref{orbifoldBC}) 
in such a way that it does not affect the symmetry breaking 
caused by the magnetic fluxes. 
Then the possible choices are 
\be
 p = \frac{2\pi n_p}{N}\brkt{1,-\frac{1}{\sqrt{3}}}, 
\ee
where $n_p=0,1,2$. 

In order for the components in (\ref{QLcomp}) 
to have zero-modes, the integers~$\kp$ and $\kp'$ in (\ref{sol:kkp}) 
should satisfy 
\bea
 \kp, \; 2\kp+\kp', \; \kp-2\kp', \; 2\kp-\kp' &\geq & 1, \nonumber\\
 -2\kp, \; -\kp+\kp', \; -2\kp-2\kp', \; -\kp-\kp' &\leq & -1, 
\eea
which are summarized as  
\be
 \kp \geq 1, \;\;\;\;\;
 -\kp+1 \leq \kp' \leq \frac{\kp-1}{2}.  \label{range:kp}
\ee
Hence, the $(\vph_0^{\sbdm{\alp}_2(k)},\vph_0^{\sbdm{\alp}_3(k)})$ 
are identified as the Higgs doublets~$H_k$  
because $k_{\sbdm{\alp}_2}=k_{\sbdm{\alp}_3}=3\kp>0$. 

The values of the orbifold twist phase~$\eta$ 
in (\ref{F:orbifoldBC}) for the relevant components are expressed as
\bea
 \eta \eql \begin{cases} \omg^{-1}e^{ip\cdot\sbdm{\alp}_2} = \omg^{n_p-1} &
 (\mbox{for $H_k$}) \\
 \omg^{-\frac{1}{2}}e^{i\vph_1}
 e^{ip\cdot\sbdm{\mu}_1} = \omg^{n_1+n_p} & 
 (\mbox{for $Q_L^j$}) \\
 \omg^{\frac{1}{2}}e^{i\vph_1}e^{ip\cdot\sbdm{\mu}_3} 
 = \omg^{n_1+1} & (\mbox{for $d_R^i$}) \\
 \omg^{\frac{1}{2}}e^{i\vph_2}e^{-ip\cdot\sbdm{\mu}_1} = \omg^{n_2+1} & 
 (\mbox{for $Q^{\prime j}_L$}) \\
 \omg^{-\frac{1}{2}}e^{i\vph_2}e^{-ip\cdot\sbdm{\mu}_3} = \omg^{n_2+n_p} & 
 (\mbox{for $u_R^i$}) \\
 \omg^{-\frac{1}{2}}e^{i\vph_3}e^{ip\cdot\sbdm{\mu}_1} = \omg^{n_3+n_p} & 
 (\mbox{for $L_L^j$}) \\ 
 \omg^{\frac{1}{2}}e^{i\vph_3}e^{ip\cdot\sbdm{\mu}_3} = \omg^{n_3+1} & 
 (\mbox{for $e_R^i$}) \\
 \omg^{\frac{1}{2}}e^{i\vph_4}e^{-ip\cdot\sbdm{\mu}_1} = \omg^{n_4+1} & 
 (\mbox{for $L^{\prime j}_L$}) \\
 \omg^{-\frac{1}{2}}e^{i\vph_4}e^{-ip\cdot\sbdm{\mu}_3} = \omg^{n_4+n_p} &
 (\mbox{for $\nu_R^i$}) \end{cases}, 
\eea
where $n_f$ ($f=1,2,3,4$) are integers (see (\ref{cstrtonvph_f})). 

From (\ref{quantize:CB}), (\ref{def:zt}), (\ref{rel:SStoWl}) and (\ref{allowedSS}), 
the allowed values of the Wilson-line phases are expressed as 
\bea
 \zeta_{\sbdm{\alp}} \eql \frac{Nc\cdot\bdm{\alp}}{2k_{\sbdm{\alp}}\pi} 
 = \frac{2}{k_{\sbdm{\alp}}}\phi^{\sbdm{\alp}}(\tau-1), 
 \nonumber\\
 \zeta_{\sbdm{\mu}f} \eql \frac{N(c\cdot\bdm{\mu}+q_fb)}{2k_{\sbdm{\mu}f}\pi} 
 = \frac{2}{k_{\sbdm{\mu}f}}\phi^{\sbdm{\mu}f}(\tau-1), 
\eea
where  
\be
 \phi^{\sbdm{\alp}} = 
 \frac{l_{\sbdm{\alp}}}{3}+\frac{1}{4}\brc{1-(-1)^{k_{\sbdm{\alp}}}}, \;\;\;\;\;
 \phi^{\sbdm{\mu}f} = 
 \frac{l_{\sbdm{\mu}f}}{3}+\frac{1}{4}\brc{1-(-1)^{k_{\sbdm{\mu}f}}}, 
 \label{possible_phis}
\ee
with $l_{\sbdm{\alp}},l_{\sbdm{\mu}f}=0,1,2$. 
Noting that $\phi^{\sbdm{\alp}_1}=0$ from the condition that 
${\rm SU(2)}_{\rm L}$ is unbroken,  
the Wilson-line phase vectors in (\ref{bg:AB_z}) should be 
\be
 \vec{c}  
 = \frac{4\pi l(\tau-1)}{3}\begin{pmatrix} 1 \\ -1/\sqrt{3} \end{pmatrix},  \;\;\;\;\;
 \vec{b} \equiv \frac{2\pi l'(\tau-1)}{3}, 
\ee
where $l$ and $l'$ are real constants. 
Then, $\phi^{\sbdm{\alp}}$ and $\phi^{\sbdm{\mu}f}$ are parametrized as 
\bea
 \phi^{\sbdm{\alp}_1} \eql 0, \;\;\;\;\;
 \phi^{\sbdm{\alp}_2} = \phi^{\sbdm{\alp}_3} = l, \nonumber\\
 \phi^{\sbdm{\mu}_11} \eql \phi^{\sbdm{\mu}_21} 
 = \frac{l}{3}, \;\;\;\;\;
 \phi^{\sbdm{\mu}_31} = -\frac{2l}{3}, \nonumber\\
 \phi^{-\sbdm{\mu}_12} \eql \phi^{-\sbdm{\mu}_22} 
 = -\frac{l}{3}+\frac{l'}{6}, \;\;\;\;\;
 \phi^{-\sbdm{\mu}_32} = \frac{2l}{3}+\frac{l'}{6}, \nonumber\\
 \phi^{\sbdm{\mu}_13} \eql \phi^{\sbdm{\mu}_23} 
 = \frac{l}{3}-\frac{l'}{3}, \;\;\;\;\;
 \phi^{\sbdm{\mu}_33} = -\frac{2l}{3}-\frac{l'}{3}, \nonumber\\
 \phi^{-\sbdm{\mu}_14} \eql \phi^{-\sbdm{\mu}_24} 
 = -\frac{l}{3}-\frac{l'}{6}, \;\;\;\;\;
 \phi^{-\sbdm{\mu}_34} = \frac{2l}{3}-\frac{l'}{6}. 
 \label{SSphases}
\eea
These phases~$\phi=\phi^{\sbdm{\alp}},\phi^{\sbdm{\mu}f}$ are defined 
modulo $\abs{K}$ ($K=k_{\sbdm{\alp}},k_{\sbdm{\mu}f}$). 
(See the comment below (\ref{BC:md_eq:2}).) 
Comparing (\ref{SSphases}) with (\ref{possible_phis}), 
we find that $2l$ ($l'$) is even for even $\kp$ ($\kp'$), 
and odd for odd $\kp$ ($\kp'$). 
\ignore{
Namely, (\ref{SSphases}) becomes 
\bea
 \phi_s^{\sbdm{\alp}_1} \eql \phi_s^{\sbdm{\alp}_2} = \phi_s^{\sbdm{\alp}_3} = 0, 
 \nonumber\\
 \phi_s^{\sbdm{\mu}_11} \eql \phi_s^{\sbdm{\mu}_21} = \phi_s^{\sbdm{\mu}_31} = 0, \nonumber\\
 \phi_s^{-\sbdm{\mu}_12} \eql \phi_s^{-\sbdm{\mu}_22} = \phi_s^{-\sbdm{\mu}_32} 
 = \frac{l}{6}, \nonumber\\
 \phi_s^{\sbdm{\mu}_13} \eql \phi_s^{\sbdm{\mu}_23} = \phi_s^{\sbdm{\mu}_33} 
 = -\frac{l}{3}, \nonumber\\
 \phi_s^{-\sbdm{\mu}_14} \eql \phi_s^{-\sbdm{\mu}_24} = \phi_s^{-\sbdm{\mu}_34} 
 = -\frac{l}{6}. 
\eea
}

In summary, the Yukawa sector of this model is specified by nine integers: 
$\kp$, $\kp'$, $l$, $l'$, $n_p$ and $n_f$~($f=1,2,3,4$). 
The numbers of zero-modes and mode functions are determined by 
the magnetic flux the field feels~$K$, 
the orbifold twist phase~$\eta$, 
and the Wilson-line phase~$\zt=\frac{2}{K}\phi(\tau-1)$, 
which are summarized in Table~\ref{parameter_for_zm}. 
\begin{table}[t]
\begin{center}
\begin{tabular}{|c||c|c|c|c|c|c|c|c|c|c|}
 \hline \rule[-2mm]{0mm}{7mm}  & $H$ & $Q_L$ & $d_R$ & 
 $Q'_L$ & $u_R$ & $L_L$ & $e_R$ & $L'_L$ & $\nu_R$ \\ \hline
 $K$ & $3\kp$ & $\kp$ & $-2\kp$ & $-\kp+\kp'$ & $2\kp+\kp'$ & $\kp-2\kp'$ & $-2\kp-2\kp'$ &
 $-\kp-\kp'$ & $2\kp-\kp'$ \\ \hline 
 $\eta$ & $\omg^{n_p-1}$ & $\omg^{n_1+n_p}$ & $\omg^{n_1+1}$ & 
 $\omg^{n_2+1}$ & $\omg^{n_2+n_p}$ & $\omg^{n_3+n_p}$ & $\omg^{n_3+1}$ & 
 $\omg^{n_4+1}$ & $\omg^{n_4+n_p}$ \\ \hline
 $\phi$ & $l$ & $\frac{l}{3}$ & $-\frac{2l}{3}$ & 
 $-\frac{2l-l'}{6}$ & $\frac{4l+l'}{6}$ & 
 $\frac{l-l'}{3}$ & $-\frac{2l+l'}{3}$ &
 $-\frac{2l+l'}{6}$ & $\frac{4l-l'}{6}$ \\ \hline 
 \end{tabular}
\end{center}
\caption{The magnetic flux~$K$ and the orbifold twist phase~$\eta$ 
felt by each field, and $\phi\equiv K\zt/2(\tau-1)$ ($\zt$: Wilson-line phase).  
The constant~$2l$ ($l'$) is even for even $\kp$ ($\kp'$), 
and odd for odd $\kp$ ($\kp'$). 
}  
\label{parameter_for_zm}
\end{table}

\subsubsection{Realization of three generations} \label{three_gen}
Here we consider the possibility that the three generations of 
quarks and leptons are realized by the magnetic fluxes. 
This occurs when $\kp=6$, $\kp'=0$, $n_p=0$, $n_{1,3}=0$, $n_{2,4}=2$ 
and $l=l'=0$.\footnote{
If we allow extra zero-modes in addition to (\ref{QLcomp}), 
other parameter choices are also possible. }
In this case, 
we obtain the following terms in the 4D effective Lagrangian from the bulk:
\bea
 \cL^{\rm (4D)} \eql -\sum_{k=1}^5\sum_{i,j=1}^3\left(y_{ij}^{(k)D}
 \bar{Q}_L^jH_kd_R^i
 +y_{ij}^{(k)U}\bar{u}_R^i\ep H_kQ_L^{\prime j} \right.\nonumber\\
 &&\left.\hspace{20mm}
 +y_{ij}^{(k)E}\bar{L}_L^jH_ke_R^i
 +y_{ij}^{(k)N}\bar{\nu}_R^i\ep H_kL_L^{\prime j}+\hc\right)+\cdots, 
 \label{cL^4D:3gen}
\eea
where $\ep H_kQ_L^{\prime j}\equiv\ep_{ab}H^a_kQ^{\prime jb}_L$ and 
$\ep H_kL_L^{\prime j}\equiv \ep_{ab}H_k^aL^{\prime jb}_L$ ($a,b$: 
${\rm SU(2)}_{\rm L}$-doublet indices), and 
\bea
 y_{ij}^{(k)D} = y_{ij}^{(k)E} 
 \eql \frac{ig}{\sqrt{2}\cdot3^{\frac{3}{4}}}\sum_{i'=1}^{12}\sum_{j'=1}^6\sum_{k'=1}^{18}
 V_{ii'}^{(\omg)}[-12,0]V_{jj'}^{(1)*}[6,0]V_{kk'}^{(\omg^2)}[18,0] \nonumber\\
 &&\times\sum_{m=1}^{18}\cF^{(-12j'-6i'-72m)}\brkt{0,-1296,0}
 \dlt_{j'-i'+6m,k'}, \nonumber\\
 y_{ij}^{(k)U} = y_{ij}^{(k)N} 
 \eql \frac{ig}{\sqrt{2}\cdot 3^{\frac{3}{4}}}\sum_{i'=1}^{12}\sum_{j'=1}^6\sum_{k'=1}^{18}
 V_{ii'}^{(\omg^2)*}[12,0]V_{jj'}^{(1)}[-6,0]V_{kk'}^{(\omg^2)}[18,0] \nonumber\\
 &&\times\sum_{m=1}^{18}\cF^{(-6i'-12j'-72m)}\brkt{0,-1296,0}
 \dlt_{i'-j'+12m,k'},  \label{Yukawa:3gen}
\eea
where $g=\bar{g}_A\simeq 0.652$ is the 4D ${\rm SU(2)}_{\rm L}$ gauge coupling, 
and we have used that~\footnote{
We can always redefine the phases of the fields 
so that the matrix elements in (\ref{matrix_element}) are real. 
}
\bea
 \langle -\bdm{\mu}_3|E_{\sbdm{\alp}_2}|-\bdm{\mu}_2\rangle \eql  
 \langle -\bdm{\mu}_3|E_{\sbdm{\alp}_3}|-\bdm{\mu}_1\rangle = -\frac{1}{\sqrt{2}}, \nonumber\\
 \langle\bdm{\mu}_2|E_{\sbdm{\alp}_2}|\bdm{\mu}_3\rangle \eql 
 \langle\bdm{\mu}_1|E_{\sbdm{\alp}_3}|\bdm{\mu}_3\rangle = \frac{1}{\sqrt{2}}. 
 \label{matrix_element}
\eea 
Extra ${\rm SU(2)}_{\rm L}$-doublets in (\ref{cL^4D:3gen}) can be made  
heavy by introducing the following brane-localized terms:
\bea
 \cL_{\rm brane} \eql \sum_{i=1}^3\left[
 \bar{\tl{Q}}_R^i(x)\brc{c_Q^i\bdm{Q}_L(x,z)
 +c_Q^{\prime i}\bdm{Q}_L^{\prime}(x,z)} \right. \nonumber\\
 &&\hspace{10mm}\left.
 +\bar{\tl{L}}_R^i(x)\brc{c_L^i\bdm{L}_L(x,z)
 +c_L^{\prime i}\bdm{L}_L^{\prime}(x,z)}+\hc\right]\dlt^{(2)}(z), 
 \label{cL_brane}
\eea
where $\tl{Q}_R^i$ and $\tl{L}_R^i$ are brane-localized 4D fields, 
and $\bdm{Q}_L$, $\bdm{Q}_L^{\prime}$, $\bdm{L}_L$ and $\bdm{L}_L^{\prime}$ 
are ${\rm SU(2)}_{\rm L}$-doublet components of 
$\Psi_-^1$, $\Psi_+^2$, $\Psi_-^3$ and $\Psi_+^4$, respectively. 
The parameters~$c_Q^i$, $c_Q^{\prime i}$, $c_L^i$ 
and $c_L^{\prime i}$ are dimensionless constants. 
Focusing on the zero-modes, (\ref{cL_brane}) is rewritten as 
\bea
 \cL_{\rm brane} \eql \sum_{i,j=1}^3\left[
 \bar{\tl{Q}}_R^i(x)\brc{m_{Q0}^{ij}Q_L^j(x)
 +m_{Q0}^{\prime ij}Q_L^{\prime j}(x)} \right.
 \nonumber\\
 &&\hspace{10mm}\left.
 +\bar{\tl{L}}_R^i(x)\brc{m_{L0}^{ij}L_L^j(x)+m_{L0}^{\prime ij}L_L^{\prime j}(x)}
 +\hc+\cdots\right]\dlt^{(2)}(z), 
\eea
where the ellipsis denotes terms involving non-zero KK modes, and 
\bea
 m_{Q0}^{ij} \defa \frac{c_Q^ih_{{\rm L}0}^{(-)\sbdm{\mu}_1 1(j)}(0)}{\sqrt{2}\pi R_1}, 
 \;\;\;\;\;
 m_{Q0}^{\prime ij} \equiv \frac{c_Q^{\prime i}h_{{\rm L}0}^{(+)-\sbdm{\mu}_1 2(j)}(0)}
 {\sqrt{2}\pi R_1}, \nonumber\\
 m_{L0}^{ij} \defa \frac{c_L^ih_{{\rm L}0}^{(-)\sbdm{\mu}_1 3(j)}(0)}{\sqrt{2}\pi R_1}, 
 \;\;\;\;\;
 m_{L0}^{ij} \equiv \frac{c_L^ih_{{\rm L}0}^{(+)-\sbdm{\mu}_1 4(j)}(0)}{\sqrt{2}\pi R_1}, 
\eea
are effective mass parameters. 
If these mass parameters are large enough, 
only the following linear combinations remain in the 4D effective theory:\footnote{
Here we neglect the mixing effect with the KK modes, which is expected to be small.  
In order to take it into account, we need to solve the modified mode equations 
that include contributions from (\ref{cL_brane}). 
}  
\be
 q_L^i \equiv V_Q^{i+3,j}Q_L^j+V_Q^{i+3,j+3}Q_L^{\prime j}, \;\;\;\;\;
 l_L^i \equiv V_L^{i+3,j}L_L^j+V_L^{i+3,j+3}L_L^{\prime j}, 
\ee
where $i=1,2,3$, and $V_Q$ and $V_L$ are $6\times 6$ unitary matrices that satisfy 
\bea
 U_Q (m_{Q0},m'_{Q0})V_Q^{-1} \eql \begin{pmatrix} \lmd_Q^1 & 0 & 0 & 0 & 0 & 0 \\
 0 & \lmd_Q^2 & 0 & 0 & 0 & 0 \\ 
 0 & 0 & \lmd_Q^3 & 0 & 0 & 0 \end{pmatrix}, \nonumber\\
 U_L (m_{L0},m'_{L0})V_L^{-1} \eql \begin{pmatrix} \lmd_L^1 & 0 & 0 & 0 & 0 & 0 \\
 0 & \lmd_L^2 & 0 & 0 & 0 & 0 \\ 
 0 & 0 & \lmd_L^3 & 0 & 0 & 0 \end{pmatrix}, 
\eea
with $3\times 3$ unitary matrices~$U_Q$ and $U_L$. 
After the extra modes are decoupled, we obtain 
\bea
 \cL^{\rm (4D)} \eql -\sum_{k=1}^5\sum_{i,j=1}^3\left(
 \tl{y}_{ij}^{(k)D}\bar{q}_L^j H_k d^i_R
 +\tl{y}_{ij}^{(k)U}\bar{u}_R^i\ep H_k q_L^j \right.\nonumber\\
 &&\hspace{20mm}\left. 
 +\tl{y}_{ij}^{(k)E}\bar{l}_L^j H_k e_R^i
 +\tl{y}_{ij}^{(k)N}\bar{\nu}_R^i\ep H_k l_L^j+\hc\right)+\cdots, 
\eea
where
\bea
 \tl{y}_{ij}^{(k)D} \defa y_{ij'}^{(k)D}(V_Q^{-1})^{j',j+3}, \;\;\;\;\;
 \tl{y}_{ij}^{(k)U} \equiv y_{ij'}^{(k)U}(V_Q^{-1})^{j'+3,j+3}, \nonumber\\
 \tl{y}_{ij}^{(k)E} \defa y_{ij'}^{(k)E}(V_L^{-1})^{j',j+3}, \;\;\;\;\;
 \tl{y}_{ij}^{(k)N} \equiv y_{ij'}^{(k)N}(V_L^{-1})^{j'+3,j+3}. 
\eea
In order to avoid large flavor-changing processes, 
we assume that only one Higgs doublet~$H_{k_0}$ acquires a nonvanishing 
vacuum expectation value (VEV). 
Then, the fermion masses are obtained as eigenvalues of 
the mass matrices given by 
\be
 M^D_{ij} = \tl{y}^{(k_0)D}_{ij}v, \;\;\;\;\;
 M^U_{ij} = \tl{y}^{(k_0)U}_{ij}v, \;\;\;\;\;
 M^E_{ij} = \tl{y}^{(k_0)E}_{ij}v, \;\;\;\;\;
 M^N_{ij} = \tl{y}^{(k_0)N}_{ij}v, 
\ee
where $v\equiv\vev{H_{k_0}}$. 
We can control the mass spectrum 
by tuning the parameters~$c_Q^i$, $c_Q^{\prime i}$, $c_L^i$ and $c_L^{\prime i}$ 
through the unitary matrices~$V_Q$ and $V_L$. 
For example, if we choose those parameters in a manner such that $V_Q\simeq\id_6$, 
we can realize the hierarchy $m_t\gg m_b$. 
In such a case, the eigenvalues of the Yukawa matrix~$\tl{y}^{(k_0)U}_{ij}$ are 
approximately given by those of $y^{(k_0)U}_{ij}$, 
whose absolute values~$\abs{\lmd^{(k_0)U}_i}$ ($i=1,2,3$) 
are shown in Appendix~\ref{3genYukawa}. 
From (\ref{lmd:Yukawa}), we find that the top quark Yukawa coupling, 
which is close to one, can be obtained when $k_0=2,5$. 
However, large hierarchies among the Yukawa couplings cannot be realized. 

Besides the Yukawa hierarchy, the existence of the five Higgs doublets 
may be problematic because 
it seems difficult to hide so many extra Higgs bosons 
from the collider experiment. 
Therefore, in the next subsection we focus on the case that only one Higgs doublet appears.

\subsubsection{One-Higgs-doublet case} \label{1HD}
Here we evaluate the magnitude of the Yukawa coupling constants 
in the case where only one Higgs doublet appears. 
This occurs when $(\kp,n_p)=(1,2),(2,0)$. 
As an example, we focus on the case~$(\kp,n_p)=(2,0)$. 
The Yukawa couplings are more restricted in the other case. 
From (\ref{range:kp}), possible values of $\kp'$ are $-1$ and $0$. 
In these cases, 
each component of (\ref{QLcomp}) has at most one zero-mode. 
Hence we will omit the ``flavor indices''~$i$ and $j$ in the following. 
The Yukawa coupling constants are expressed as follows:
\begin{description}
\item[(i) $\bdm{\kp'=0}$ case] 
\bea
 y^D \eql Y^{(-)}\brkt{n_1,-\frac{2l}{3},\frac{l}{3}}, \;\;\;\;\;
 y^U = Y^{(+)}\brkt{n_2,\frac{4l+l'}{6},-\frac{2l-l'}{6}}, \nonumber\\
 y^E \eql Y^{(-)}\brkt{n_3,-\frac{2l+l'}{3},\frac{l-l'}{3}}, \;\;\;\;\;
 y^N = Y^{(+)}\brkt{n_4,\frac{4l-l'}{6},-\frac{2l+l'}{6}}, 
\eea
where $l$ is an integer, $l'$ is an even number, and 
\bea
 Y^{(+)}(n,\phi_1,\phi_2) \defa 
 \frac{ig}{\sqrt{2}\cdot 3^{\frac{1}{4}}}\sum_{i'=1}^4\sum_{j'=1}^2\sum_{k'=1}^6
 V_{1i'}^{(\omg^n)*}[4,\phi_1]V_{1j'}^{(\omg^{n+1})}[-2,\phi_2]
 V_{1k'}^{(\omg^{-1})}\sbk{6,\phi_1-\phi_2} \nonumber\\
 &&\times\sum_{m=1}^6\cF^{(-2i'-4j'-8m)}\brkt{0,-48,\frac{(\phi_1+2\phi_2)(\tau-1)}{12}}
 \dlt_{i'-j'+4m,k'}, \nonumber\\
 Y^{(-)}(n,\phi_1,\phi_2) \defa 
 \frac{ig}{\sqrt{2}\cdot 3^{\frac{1}{4}}}\sum_{i'=1}^4\sum_{j'=1}^2\sum_{k'=1}^6
 V_{1i'}^{(\omg^{n+1})}[-4,\phi_1]V_{1j'}^{(\omg^n)*}[2,\phi_2]
 V_{1k'}^{(\omg^{-1})}[6,\phi_2-\phi_1] \nonumber\\
 &&\times\sum_{m=1}^6\cF^{(-2i'-4j'-8m)}\brkt{0,-48,\frac{(\phi_1+2\phi_2)(\tau-1)}{12}}
 \dlt_{j'-i'+2m,k'}, \nonumber\\ \label{expr:Y^pm}
\eea
where $\phi_a$ ($a=1,2$) are defined by $\zt_a=\frac{2\phi_a}{K_a}(\tau-1)$, 
and here we choose them as the second argument of $V_{ij}^{(\eta)}$ instead of $\zeta_a$. 
The possible values of $n$, $\phi_1$ and $\phi_2$ in (\ref{expr:Y^pm}) are 
\bea
%
%
 n \eql 0, \; 1, \; 2, \;\;\;\;\; \mbox{(mod 3)} \nonumber\\
 \phi_1 \eql \phi_2-{\rm floor}(\phi_2)+u, \;\;\;\;\; \mbox{(mod 4)} \nonumber\\
 \phi_2 \eql 0, \; \frac{1}{3}, \; \frac{2}{3}, \; 1, \; \frac{4}{3}, \; \frac{5}{3}, 
 \;\;\;\;\; \mbox{(mod $2$)} 
\eea
where $u=0,1,2,3$. 
Numerical values of $\abs{Y^{(\pm)}}$ are listed in Table~\ref{k0case} 
of Appendix~\ref{num_values}. 
From the table, we can see that possible values of the Yukawa coupling constants are 
\be
 \abs{y^{D,U,E,N}} = 0.191, \; 0.270, \; 0.369, \; 0.522, \; 0.573, \; 0.811. 
\ee

\item[(ii) $\bdm{\kp'=-1}$ case]
\bea
 y^D \eql Y^{(-)}\brkt{n_1,-\frac{2l}{3},\frac{l}{3}}, \nonumber\\
 y^U \eql \frac{ig}{\sqrt{2}\cdot 3^{\frac{1}{4}}}\sum_{i'=1}^3\sum_{j'=1}^3\sum_{k'=1}^6
 V_{1i'}^{(\omg^{n_2})*}\sbk{3,\frac{4l+l'}{6}}
 V_{1j'}^{(\omg^{n_2+1})}\sbk{-3,-\frac{2l-l'}{6}}
 V_{1k'}^{(\omg^{-1})}\sbk{6,l} \nonumber\\
 &&\times\sum_{m=1}^6\cF^{(-3i'-3j'-9m)}\brkt{0,-54,-\frac{2(l+l')(\tau-1)}{54}}
 \dlt_{i'-j'+3m,k'}, \nonumber\\
 y^E \eql 
 Y^{(+)}\brkt{n_3,\frac{l-l'}{3},-\frac{2l+l'}{3}}, \nonumber\\
 y^N \eql \frac{ig}{\sqrt{2}\cdot 3^{\frac{1}{4}}}\sum_{i'=1}^5\sum_{k'=1}^6
 V_{1i'}^{(\omg^{n_4})*}\sbk{5,\frac{4l-l'}{6}}
 V_{11}^{(\omg^{n_4+1})}\sbk{-1,-\frac{2l+l'}{6}}
 V_{1k'}^{(\omg^{-1})}\sbk{6,l} \nonumber\\
 &&\times\sum_{m=1}^6\cF^{(-i'-5-5m)}\brkt{0,-30,-\frac{(l-l')(\tau-1)}{45}}
 \dlt_{i'-1+5m,k'}.
\eea
Numerical values of these are summarized in Tables~\ref{k1case:U} and \ref{k1case:N} 
of Appendix~\ref{num_values}. 
From the tables, we can see that the Yukawa coupling constants take the following values: 
\bea
 \abs{y^{D,E}} \eql 0.191, \; 0.270, \; 0.369, \; 0.522, \; 0.573, \; 0.811, \nonumber\\
 \abs{y^U} \eql 0.365, \; 0.430, \; 0.461, \; 0.667, \; 0.798,  \nonumber\\
 \abs{y^N} \eql 0.101, \; 0.176, \; 0.188, \; 0.288, \; 0.533, \; 
 0.541, \; 0.559, \; 0.924. 
\eea
\end{description}

In each case of Sec.~\ref{three_gen} and Sec.~\ref{1HD}, 
the eigenvalues of the Yukawa matrices are within the region~$[0.1,1]$, 
and we cannot realize small Yukawa couplings only by means of 
the magnetic fluxes and the Wilson-line phases. 
We need an additional mechanism to obtain them. 
This is mainly due to the matrices~$V_{ij}^{(\eta)}$ 
in (\ref{expr:y_ijk}) and (\ref{expr:y_ijk:2}). 
In order to see this, let us define the quantity:
\be
 \tl{y}^{(k)}_{ij}(\kp) \equiv \frac{ig}{3^{\frac{1}{4}}\sqrt{\kp}}
 \sum_{m=1}^{3\kp}\cF^{(-\kp i-2\kp j-2\kp^2 m)}\brkt{0,-6\kp^3,0}
 \dlt_{i-j+2\kp m,k}, 
 \label{tly_ijk}
\ee
which is obtained from (\ref{expr:y_ijk}) in the case of our model 
by taking $\kp'=l'=0$ 
and replacing the $V_{ij}^{(\eta)}$ matrices with $\dlt_{ij}$. 
The indices~$i$ and $j$ are assumed to run from 1 to $\kp$. 
Then, we can see that 
the eigenvalues of (\ref{tly_ijk}), $\tl{\lmd}_i^{(k)}(\kp)$ ($i=1,\cdots,\kp$), 
can take small values. 
For example, $|\lmd^{(k)}_i(2)|$ take values in the range~$[6.04\times 10^{-4},0.843]$, 
and $|\lmd^{(k)}_i(4)|$ are in $[2.05\times 10^{-7},1.12]$. 

We should also note that the top quark Yukawa coupling, which is close to 1, 
can be reproduced in our model, which only has the small representations~$\bdm{3}$ 
and $\bdm{\bar{3}}$. 
This is in contrast to a model without the magnetic fluxes. 
In the absence of the magnetic fluxes, the zero-mode wave functions are constants 
unless the brane-localized terms exist. 
In such a case, the Yukawa couplings are equal to $1/\sqrt{2}$. 
Thus, we need an enhancement factor, which is roughly $\sqrt{2}$, 
in order to obtain the top quark mass.  
This can be accomplished by embedding the quark fields 
into a larger representation of SU(3). 
In the presence of the magnetic fluxes, on the other hand, 
such an enhancement factor is obtained 
as an overlap integral of the mode functions that have nontrivial profiles.

\section{Summary} \label{summary}
We have studied the Yukawa couplings in 6D gauge-Higgs unification models 
compactified on an orbifold~$T^2/Z_N$ 
in the presence of background magnetic fluxes. 
The effects of the magnetic fluxes are multiplication of zero-modes for each 6D field 
and deformation of the constant mode functions for the zero-modes. 
The former opens up the interesting possibility that the generational structure 
of quarks and leptons is realized, and the latter is essential to controlling the 
magnitude of the Yukawa coupling constants. 

We considered a $G\times{\rm U(1)}_X$ gauge theory, where $G$ is a simple group, 
and introduced the magnetic fluxes for ${\rm U(1)}_X$ and the Cartan part of $G$. 
The number of zero-modes are determined by the orbifold boundary conditions, 
and the fluxes and the Wilson-line phases that the 6D field actually feels. 
It should be emphasized that all these quantities are quantized. 
Thus the Yukawa sector is controlled by a finite number of integers. 
As a specific model, we consider an ${\rm SU(3)}\times{\rm U(1)}_X$ gauge theory 
on $T^2/Z_3$ with four 6D Weyl fermions belonging to $\bdm{3}$ or $\bdm{\bar{3}}$. 
We evaluated the Yukawa coupling constants 
in cases where three generations are realized, 
and where only one Higgs doublet appears in the 4D effective theory. 
The Yukawa sector of our model is specified by nine integers. 
Due to this property and the symmetric structure of the Yukawa coupling formula, 
the coupling constants can take only limited numbers of values. 
They are all within the region~$[0.1,1]$. 
This stems from the fact that the mode functions on $T^2/Z_3$ 
are given by mixtures of those on $T^2$. 
This mixing effect makes the profiles of the mode functions complicated. 
Thus it is difficult to realize the observed large hierarchy among the fermion masses 
only by means of the magnetic fluxes and the Wilson-line phases. 
We need additional mechanism to obtain it. 
The situation is similar in models on $T^2/Z_4$ or $T^2/Z_6$. 
In the case of $T^2/Z_2$, the mixing matrices~$V_{ij}^{(\eta)}$ in (\ref{expr:y_ijk}) 
and (\ref{expr:y_ijk:2}) become diagonal, 
and thus small Yukawa couplings can easily be obtained~\cite{Abe:2015yva}.  
We should also note that there is an advantageous feature 
of a model with magnetic fluxes. 
We can realize the top quark Yukawa coupling 
without introducing a large representation of $G$, 
thanks to the nontrivial profiles of the zero-mode wave functions. 


In this work, we neglected the mixing with the KK modes 
induced by the brane-localized terms and the Higgs VEVs. 
Such effects are important in evaluating  
the deviation of each coupling constant from the standard model value. 
They can be taken into account by solving the mode equations 
in the presence of the brane-localized terms and the $W_z^{\sbdm{\alp}}$ background. 
This will be discussed in a subsequent paper.

\subsection*{Acknowledgements}
The authors would like to thank Yoshiyuki Tatsuta for valuable information. 
This work was supported in part by 
Grant-in-Aid for Scientific Research (C) No.~25400283  
from Japan Society for the Promotion of Science (Y.S.).

\appendix

\section{Analytic forms of $\bdm{D_{jk}^{(\omg^l)}}$ in (\ref{def:C_jk})} \label{C:anl_fm}
Here we collect the analytic forms of $D_{jk}^{(\omg^l)}$ in (\ref{def:C_jk}) 
obtained in Ref.~\cite{Abe:2014noa}. 
In the following formulae, we choose a gauge in which 
the Wilson-line phases are zero. 
The correspondence to the Wilson-line phases in the text can be read off 
from (\ref{rel:WltoSS}) or (\ref{rel:SStoWl}). 
Here, $K$, $\phi_1$ and $\phi_\tau$ collectively denote 
$\{k_{\sbdm{\alp}},k_{\sbdm{\mu}f}\}$, $\{\phi_1^{\sbdm{\alp}},\phi_1^{\sbdm{\mu}f}\}$ 
and $\{\phi_\tau^{\sbdm{\alp}},\phi_\tau^{\sbdm{\mu}f}\}$, respectively. 
The SS phases can only take discrete values on $T^2/Z_N$ 
from the consistency conditions~\cite{Abe:2013bca}. 
This is equivalent to only discrete values of the Wilson-line phases 
being allowed~\cite{Ibanez:1986tp,Kobayashi:1990mi,Abe:2009uz}. 

Note that $D_{jk}^{(1)}=\dlt_{jk}$ by definition. 
The other coefficients~$D_{jk}^{(\omg^l)}$ ($l\neq 0$) are shown in the following. 

\begin{description}
\item[$\bdm{T^2/Z_2}$] \mbox{}\\
The allowed values of the SS phases are
\be
 (\phi_1,\phi_\tau) = (0,0), \; \brkt{\frac{1}{2},0}, \;
 \brkt{0,\frac{1}{2}}, \; \brkt{\frac{1}{2},\frac{1}{2}}. 
\ee
The explicit form of $D_{jk}^{(-1)}$ is 
\bea
 D_{jk}^{(-1)} \eql \exp\brc{-\frac{4\pi i}{K}\phi_\tau(\phi_1+j)}\dlt_{-2\phi_1-j,k} 
 \nonumber\\
 \eql \exp\brc{\frac{4\pi i}{K}\phi_\tau(\phi_1+k)}\dlt_{-2\phi_1-k,j}
 = \brc{D^{(-1)\dagger}}_{jk}. 
\eea

\item[$\bdm{T^2/Z_3}$] \mbox{}\\
The allowed values of the SS phases are
\be
 \phi \equiv \phi_1 = \phi_\tau = \begin{cases} \displaystyle
 0, \; \frac{1}{3}, \; \frac{2}{3} & (\mbox{$K$: even}) \\[4mm] \displaystyle 
 \frac{1}{6}, \; \frac{1}{2}, \; \frac{5}{6} & (\mbox{$K$: odd}) \end{cases}. 
 \label{allowedSS}
\ee
The explicit forms of $D_{jk}^{(\omg^l)}$ are 
\bea
 D_{jk}^{(\omg)} \eql \frac{e^{-\sgn(K)\frac{\pi i}{12}}}{\sqrt{\abs{K}}}\exp\brc{
 \frac{\pi i}{K}\brkt{3\phi^2+k(k+6\phi)+2jk}}, \nonumber\\
 D_{jk}^{(\omg^2)} \eql \frac{e^{\sgn(K)\frac{\pi i}{12}}}{\sqrt{\abs{K}}}\exp\brc{
 -\frac{\pi i}{K}\brkt{3\phi^2+j(j+6\phi)+2jk}} 
 = \brc{D^{(\omg)\dagger}}_{jk}. 
 \label{expr:C_jk:Z3}
\eea

\item[$\bdm{T^2/Z_4}$] \mbox{}\\
The allowed values of the SS phases are
\be
 \phi \equiv \phi_1 = \phi_\tau = 0, \; \frac{1}{2}. 
\ee
The explicit forms of $D_{jk}^{(\omg^l)}$ are 
\bea
 D_{jk}^{(\omg)} \eql \frac{1}{\sqrt{\abs{K}}}
 \exp\brc{\frac{2\pi i}{K}\brkt{\phi^2+2\phi k+jk}}, \nonumber\\
 D_{jk}^{(\omg^2)} \eql \exp\brc{-\frac{4\pi i}{K}\phi(\phi+j)}\dlt_{-2\phi-j,k} 
 = \brc{D^{(\omg^2)\dagger}}_{jk}, 
 \nonumber\\
 D_{jk}^{(\omg^3)} \eql \frac{1}{\sqrt{\abs{K}}}
 \exp\brc{-\frac{2\pi i}{K}\brkt{\phi^2+\phi j+jk}}
 = \brc{D^{(\omg)\dagger}}_{jk}. 
\eea

\item[$\bdm{T^2/Z_6}$] \mbox{}\\
The allowed values of the SS phases are 
\be
 \phi \equiv \phi_1 = \phi_\tau = \begin{cases} \displaystyle 0 & (\mbox{$K$: even}) \\
 \displaystyle \frac{1}{2} & (\mbox{$K$: odd}) \end{cases}. 
\ee
The explicit forms of $D_{jk}^{(\omg^l)}$ are 
\bea
 D_{jk}^{(\omg)} \eql \frac{e^{\sgn(K)\frac{\pi i}{12}}}{\sqrt{\abs{K}}}\exp\brc{
 \frac{\pi i}{K}\brkt{\phi^2-k(k-2\phi)+2jk}}, \nonumber\\
 D_{jk}^{(\omg^2)} \eql \frac{e^{-\sgn(K)\frac{\pi i}{12}}}{\sqrt{\abs{K}}}\exp\brc{
 \frac{\pi i}{K}\brkt{3\phi^2+j(j+2\phi)+2k(j+2\phi)}}, \nonumber\\
 D_{jk}^{(\omg^3)} \eql \exp\brc{-\frac{4\pi i}{K}\phi(\phi+j)}\dlt_{-2\phi-j,k}
 = \brc{D^{(\omg^3)\dagger}}_{jk}, 
 \nonumber\\
 D_{jk}^{(\omg^4)} \eql \frac{e^{\sgn(K)\frac{\pi i}{12}}}{\sqrt{\abs{K}}}\exp\brc{
 -\frac{\pi i}{K}\brkt{3\phi^2+k(k+2\phi)+2j(k+2\phi)}} 
 = \brc{D^{(\omg^2)\dagger}}_{jk}, \nonumber\\
 D_{jk}^{(\omg^5)} \eql \frac{e^{-\sgn(K)\frac{\pi i}{12}}}{\sqrt{\abs{K}}}\exp\brc{
 -\frac{\pi i}{K}\brkt{\phi^2-j(j-2\phi)+2jk}} 
 = \brc{D^{(\omg)\dagger}}_{jk}. 
\eea

The sign function~$\sgn(K)$ comes from the formula:
\be
 \sum_{s=0}^{\abs{K}-1}\exp\brc{\frac{\pi i}{K}(s+\bt)^2} 
 = \sqrt{\abs{K}}e^{\sgn(K)\frac{\pi i}{4}}, 
\ee
where $\bt$ is an integer (half-integer) when $K$ is even (odd). 

\end{description}

\section{Normalizations of KK modes}
In this appendix, we identify the coefficients in (\ref{KKexpand}). 
Here we focus on those for $W_\mu^{\sbdm{\alp}}$ and $W_z^{\sbdm{\alp}}$. 
The other normalization factors are obtained similarly. 
The 6D Lagrangian~(\ref{6DcL}) includes the following terms: 
\bea
 \cL \eql -\frac{1}{4g_A^2}\tr\brc{F^{\mu\nu}F_{\mu\nu}
 +\frac{2}{(\pi R_1)^2}F^{\mu\bar{z}}F_{\mu z}}+\cdots \nonumber\\
 \eql -\frac{1}{4g_A^2}\sum_{\sbdm{\alp}}\brc{
 \brkt{W^{\sbdm{\alp}\mu\nu}}^*W_{\mu\nu}^{\sbdm{\alp}}
 +\frac{2}{(\pi R_1)^2}\brkt{W^{\sbdm{\alp}\mu z}}^*W_{\mu z}^{\sbdm{\alp}}}+\cdots, 
 \label{cL:6D}
\eea
where 
\bea
 W_{\mu M}^{\sbdm{\alp}} \defa \der_\mu W_M^{\sbdm{\alp}}-\der_M W_\mu^{\sbdm{\alp}}
 -i\left\{\sum_i\alp^i\brkt{C_\mu^iW_M^{\sbdm{\alp}}-W_\mu^{\sbdm{\alp}}C_M^i}
 +\sum_{\sbdm{\bt}}N_{\sbdm{\bt},\sbdm{\alp}-\sbdm{\bt}}
 W_\mu^{\sbdm{\bt}}W_M^{\sbdm{\alp}-\sbdm{\bt}}\right\}, \nonumber\\
 N_{\sbdm{\bt},\sbdm{\gm}} \defa \langle\bdm{\bt}+\bdm{\gm}|E_{\sbdm{\bt}}|\bdm{\gm}\rangle. 
\eea
The KK expansion is expressed as
\bea
 W_\mu^{\sbdm{\alp}}(x,z) \eql \cN_W\sum_n f_n^{\sbdm{\alp}}(z)W_\mu^{\sbdm{\alp}(n)}(x), 
 \nonumber\\
 W_z^{\sbdm{\alp}}(x,z) \eql \cN_\vph\sum_n g_n^{\sbdm{\alp}}(z)\vph_n^{\sbdm{\alp}}(x), 
 \label{KKexpand:W}
\eea
where $\cN_W$ and $\cN_\vph$ are positive constants, and 
the mode functions are normalized as
\be
 \int_{T^2/Z_N}\dr^2z\;\brc{f_n^{\sbdm{\alp}}(z)}^*f_m^{\sbdm{\alp}}(z) 
 = \int_{T^2/Z_N}\dr^2z\;\brc{g_n^{\sbdm{\alp}}(z)}^*g_m^{\sbdm{\alp}}(z) = \dlt_{nm}. 
\ee
Substituting (\ref{KKexpand:W}) into (\ref{cL:6D}), 
we obtain the 4D effective Lagrangian: 
\bea
 \cL^{\rm (4D)} \eql \int_{T^2/Z_N}dx^4dx^5\;\cL = 2(\pi R_1)^2\int_{T^2/Z_N}d^2z\;\cL 
 \nonumber\\
 \eql -\frac{2(\pi R_1)^2}{4g_A^2}\int_{T^2/Z_N}\left\{
 2\abs{W_{\mu\nu}^{\sbdm{\alp}_1}}^2
 +\frac{2}{(\pi R_1)^2}\left(
 \abs{\der_\mu W_z^{\sbdm{\alp}_2}-iN_{-\sbdm{\alp}_1,\sbdm{\alp}_3}
 W_\mu^{-\sbdm{\alp}_1}W_z^{\sbdm{\alp}_3}+\cdots}^2
 \right.\right.\nonumber\\
 &&\hspace{50mm}\left.\left.
 +\abs{\der_\mu W_z^{\sbdm{\alp}_3}-iN_{\sbdm{\alp}_1,\sbdm{\alp}_2}W_\mu^{\sbdm{\alp}_1}
 W_z^{\sbdm{\alp}_2}+\cdots}^2
 \right)\right\}+\cdots \nonumber\\
 \eql -\frac{\cN_W^2(\pi R_1)^2}{g_A^2}
 \abs{\der_\mu W_\nu^{\sbdm{\alp}_1(0)}-\der_\nu W_\mu^{\sbdm{\alp}_1(0)}}^2
 -\frac{\cN_\vph^2}{g_A^2}\brkt{\abs{\cD_\mu\vph_0^{\sbdm{\alp}_2}}^2
 +\abs{\cD_\mu\vph_0^{\sbdm{\alp}_3}}^2} \nonumber\\
 &&+\cdots,  \label{cL:4D}
\eea
where $\bdm{\alp}_1$ and $\{\bdm{\alp}_2,\bdm{\alp}_3\}$ are 
the roots such that $W_\mu^{\pm\sbdm{\alp}_1}$ and $W_z^{\sbdm{\alp}_{2,3}}$ have 
zero-modes that are identified with the W boson and the Higgs doublet fields respectively, and 
\bea
 \cD_\mu\vph_0^{\sbdm{\alp}_2} \defa 
 \der_\mu\vph_0^{\sbdm{\alp}_2}-iN_{-\sbdm{\alp}_1,\sbdm{\alp}_3}
 \cN_Wf_0^{-\sbdm{\alp}_1}(z)W_\mu^{-\sbdm{\alp}_1(0)}\vph_0^{\sbdm{\alp}_3} \nonumber\\
 \cD_\mu\vph_0^{\sbdm{\alp}_3} \defa 
 \der_\mu\vph_0^{\sbdm{\alp}_3}
 -iN_{\sbdm{\alp}_1,\sbdm{\alp}_2}
 \cN_Wf_0^{\sbdm{\alp}_1}(z)W_\mu^{\sbdm{\alp}_1(0)}\vph_0^{\sbdm{\alp}_2}. 
 \label{def:cD_muvph}
\eea
We have used that 
\be
 f_0^{\pm\sbdm{\alp}_1}(z) = \sqrt{\frac{N}{2\Im\tau}}. 
\ee

Comparing (\ref{cL:4D}) with the standard model, 
\bea
 \cL_{\rm SM} \eql -\frac{1}{2}\tr\brc{\brkt{\sum_aF^a_{\mu\nu}\frac{\sgm^a}{2}}^2}
 -\abs{\brkt{\der_\mu-igA^a_\mu\frac{\sgm^a}{2}}
 \begin{pmatrix} \vph^+ \\ \vph^0 \end{pmatrix}}^2+\cdots \nonumber\\
 \eql -\frac{1}{4}\brkt{\abs{F^1_{\mu\nu}}^2+\abs{F^2_{\mu\nu}}^2+\cdots} 
 -\abs{\der_\mu\vph^0-ig(A_\mu^1+iA_\mu^2)\vph^+}^2 \nonumber\\
 &&-\abs{\der_\mu\vph^+-ig(A_\mu^1-iA_\mu^2)\vph^0}^2+\cdots \nonumber\\
 \eql -\frac{1}{2}\abs{\der_\mu W_\nu^+-\der_\nu W_\mu^+}^2
 -\abs{\der_\mu\vph^0-\frac{ig}{\sqrt{2}}W^-_\mu\vph^+}^2 
 -\abs{\der_\mu\vph^+-\frac{ig}{\sqrt{2}}W^+_\mu\vph^0}^2 \nonumber\\
 &&+\cdots, 
\eea
where $A_\mu^a$ ($a=1,2,3$) are the ${\rm SU(2)}_{\rm L}$ gauge fields, 
$F^a_{\mu\nu}$ are their field strengths, and 
$W^\pm_\mu\equiv\frac{1}{\sqrt{2}}(A_\mu^1\mp iA_\mu^2)$, 
the constants~$\cN_W$ and $\cN_\vph$ should be chosen as
\be
 \frac{\cN_W^2(\pi R_1)^2}{g_A^2} = \frac{1}{2}, \;\;\;\;\;
 \frac{\cN_\vph^2}{g_A^2} = 1, 
\ee
and the 4D gauge coupling constant~$\bar{g}_A$ is identified from (\ref{def:cD_muvph}) as
\be
 \frac{\bar{g}_A}{\sqrt{2}} = N_{-\sbdm{\alp}_1,\sbdm{\alp}_3}\cN_W\sqrt{\frac{N}{2\Im\tau}}
 = N_{\sbdm{\alp}_1,\sbdm{\alp}_2}\cN_W\sqrt{\frac{N}{2\Im\tau}}. 
\ee
Solving these, we obtain 
\be
 \cN_W = \frac{g_A}{\sqrt{2}\pi R_1}, \;\;\;\;\;
 \cN_\vph = g_A, \;\;\;\;\;
 \bar{g}_A = \frac{g_A}{\sqrt{\cA}}. 
\ee
We have used that 
\be
 N_{-\sbdm{\alp_1},\sbdm{\alp}_3} = N_{\sbdm{\alp}_1,\sbdm{\alp}_2} = \frac{1}{\sqrt{2}}. 
\ee
after appropriate phase redefinitions of the fields.

\ignore{
\section{Operator analysis}
In this appendix, we review the operator analysis in Ref.~\cite{Abe:2014noa}. 
We focus on the fields that feel the magnetic fluxes. 
Here we choose a gauge where the Wilson-line phases are zero. 
Then, $(k_{\sbdm{\alp}},k_{\sbdm{\mu}f})$ and 
$(D_z^{(\sbdm{\alp})},D_z^{(\sbdm{\mu}f)})$, 
$(D_{\bar{z}}^{(\sbdm{\alp})},D_{\bar{z}}^{(\sbdm{\mu}f)})$ 
are collectively denoted by $K$ and 
\be
 D_z = \der_z-c{\bar{z}}, \;\;\;\;\;
 D_{\bar{z}} = \der_{\bar{z}}+cz, 
\ee
where $c\equiv\pi K/2\Im\tau$, respectively. 
\subsection{Torus case}
Note that 
\bea
 D_zD_{\bar{z}} \eql -\frac{1}{4}\vec{\cP}^2+c, \;\;\;\;\;
 \mbox{(when $K>0$)} \nonumber\\
 D_{\bar{z}}D_z \eql -\frac{1}{4}\vec{\cP}^2+\abs{c}, \;\;\;\;\;
 \mbox{(when $K<0$)}
\eea
where
\bea
 \vec{\cP} \defa \begin{pmatrix} -i\der_1+2cy_2 \\ -i\der_2-2cy_1 
 \end{pmatrix} = -i\vec{\nabla}+\Omg\vec{y}, \nonumber\\
 \vec{y} \eql \begin{pmatrix} y_1 \\ y_2 \end{pmatrix} 
 \equiv \frac{1}{2\pi R_1}\begin{pmatrix} x^4 \\ x^5 \end{pmatrix}, 
 \;\;\;\;\;
 \Omg \equiv \begin{pmatrix} 0 & 2c \\ -2c & 0 \end{pmatrix}. 
\eea
Thus, the mode equations in (\ref{md_eq}) are collectively expressed as
\be
 \vec{\cP}^2\tl{F} = 4\brkt{\abs{\tl{m}_n}^2+\abs{c}}\tl{F}, 
 \label{md_eq:2}
\ee
where $\tl{F}(z)$ is a mode function on $T^2$. 
The torus boundary conditions in (\ref{BC:md_eq:1}) are collectively expressed as
\bea
 \tl{F}(\vec{y}+\vec{u}_a) = \exp\brc{i\brkt{
 \vec{u}_a^t\Omg\vec{y}+2\pi\phi_a}}\tl{F}(\vec{y}), 
\eea
where $a=1,2$, and 
\be
 \vec{u}_1 \equiv \begin{pmatrix} 1 \\ 0 \end{pmatrix}, 
 \;\;\;\;\;
 \vec{u}_2 \equiv \begin{pmatrix} \Re\tau \\ \Im\tau \end{pmatrix}, 
\ee
and $\phi_1$ and $\phi_2\equiv\phi_\tau$ are the SS phases. 
This can be reexpressed as 
\be
 \exp\brc{i\vec{u}_a^t\brkt{\vec{\cP}-2\Omg\vec{y}}-2\pi i\phi_a}\tl{F}(\vec{y})
 = \tl{F}(\vec{y}). 
 \label{periodic_cond:1}
\ee
Here we define operators:
\bea
 X \defa -\frac{1}{2\sqrt{\abs{c}}}\cP_2, \;\;\;\;\;
 P \equiv \frac{1}{2\sqrt{\abs{c}}}\cP_1, \nonumber\\
 \tl{X} \defa \frac{1}{2\pi K}\brc{
 \vec{u}_1^t\brkt{\vec{\cP}-2\Omg\vec{y}}-2\pi\phi_1}, \;\;\;\;\;
 \tl{P} \equiv -\brc{
 \vec{u}_2^t\brkt{\vec{\cP}-2\Omg\vec{y}}-2\phi\phi_2}. 
\eea
Then, they satisfy 
\be
 \sbk{X,P} = i, \;\;\;\;\;
 \sbk{\tl{X},\tl{P}} = i, 
\ee
and the other commutators vanish. 
In terms of these operators, $\vec{\cP}^2$ in (\ref{md_eq:2}) 
and (\ref{periodic_cond:1}) are rewritten as
\be
 H \equiv \vec{\cP}^2 = 4\abs{c}\brkt{X^2+P^2}, 
 \label{def:H}
\ee
and 
\be
 e^{2\pi Ki\tl{X}}\tl{F}(\vec{y}) = \tl{F}(\vec{y}), \;\;\;\;\;
 e^{-i\tl{P}}\tl{F}(\vec{y}) = \tl{F}(\vec{y}). \label{periodic_cond:2}
\ee
The operator~$H$ can be regarded as the Hamiltonian of 
the one-dimensional harmonic oscillator. 
Therefore, it is rewritten as 
\be
 H = 4\abs{c}\brkt{2\hat{a}^\dagger\hat{a}+1}, 
\ee
where
\be
 \hat{a} \equiv \frac{1}{\sqrt{2}}\brkt{X+iP}, 
\ee
and the eigenstates are expressed as
\be
 |n\rangle = \frac{1}{\sqrt{n!}}(\hat{a}^\dagger)^n|0\rangle, 
\ee
where $|0\rangle$ is the lowest state that satisfies 
$\hat{a}|0\rangle=0$, and $n=0,1,2,\cdots$. 
The corresponding eigenvalues are 
\be
 4\brkt{\abs{\tl{m}_n}^2+\abs{c}} = 8\abs{c}\brkt{n+\frac{1}{2}}, 
\ee
which indicates that
\be
 \abs{\tl{m}_n}^2 = 2\abs{c}n = \frac{\pi\abs{K}n}{\Im\tau}. 
\ee
Since $\tl{X}$ commutes with $H$, we can take simultaneous eigenstates 
of them as a basis of the ``state vector space''. 
Namely, the mode function~$\tl{F}(z)$ can be denoted by
\be
 \tl{F}(z) = \left|n,\frac{j}{K}\right\rangle, 
\ee
where $j/K$ is an eigenvalue of $\tl{X}$. 
Then (\ref{periodic_cond:2}) indicates that $j$ is an integer. 
The commutation relation between $\tl{X}$ and $\tl{P}$ suggest that
$\tl{P}$ is a translation operator for the ``coordinate''~$j$. 
In fact, since we can show that
\be
 \tl{X}\brkt{e^{-ib\tl{P}}\left|n,\frac{j}{K}\right\rangle} 
 = \brkt{\frac{j}{K}+b}e^{-ib\tl{P}}\left|n,\frac{j}{K}\right\rangle, 
\ee
where $b$ is a real constant, we find that
\be
 e^{-ib\tl{P}}\left|n,\frac{j}{K}\right\rangle 
 = \left|n,\frac{j}{K}+b\right\rangle. 
\ee
Thus, the first condition in (\ref{periodic_cond:2}) indicates that
$j$ is defined on $Z_{\abs{K}}$, \ie, 
\be
 j = 0,1,\cdots,\abs{K}-1. 
\ee
The orthonormal condition is
\be
 \left\langle n,\frac{j}{K}\right|\left.l,\frac{k}{K}\right\rangle
 = \dlt_{n,l}\dlt_{j,k}. 
\ee
\subsection{Orbifold case}
Here we consider a case of $T^2/Z_3$ as an example. 
Under the $Z_3$ rotation, 
\be
 \vec{y} \to U_\omg\vec{y}U_\omg^{-1} = R_\omg\vec{y}, 
\ee
where
\be
 R_\omg \begin{pmatrix} -\frac{1}{2} & -\frac{\sqrt{3}}{2} \\
 \frac{\sqrt{3}}{2} & -\frac{1}{2} \end{pmatrix}. 
\ee
Thus, 
\bea
 U_\omg\tl{X}U_\omg^{-1} \eql -\tl{X}+\frac{\tl{P}}{2\pi K}
 -\frac{1}{K}\brkt{2\phi_1+\phi_2}, \nonumber\\
 U_\omg\tl{P}U_\omg^{-1} \eql -2\pi K\tl{X}-2\pi\brkt{\phi_1-\phi_2},  
\eea
since 
\be
 \vec{u}_1^tR_\omg = -\vec{u}_1-\vec{u}_2, \;\;\;\;\;
 \vec{u}_2^tR_\omg = \vec{u}_1, \;\;\;\;\;
 \Omg R_\omg = R_\omg\Omg. 
\ee
These can be rewritten as
\bea
 U_\omg^{-1}\tl{X}U_\omg \eql -\frac{\tl{P}}{2\pi K}
 -\frac{\phi_1-\phi_2}{K}, \nonumber\\
 U_\omg^{-1}\tl{P}U_\omg \eql 2\pi K\tl{X}-\tl{P}
 +2\pi\brkt{\phi_1+2\phi_2}. 
 \label{UXU:2}
\eea
\subsubsection{Allowed Scherk-Schwarz phases}
In the bra-ket notation, (\ref{periodic_cond:2}) is expressed as
\be
 e^{2\pi Ki\tl{X}}\left|n,\frac{j}{K}\right\rangle 
 = \left|n,\frac{j}{K}\right\rangle, \;\;\;\;\;
 e^{i\tl{P}}\left|n,\frac{j}{K}\right\rangle 
 = \left|n,\frac{j}{K}\right\rangle. 
\ee
Since a $Z_3$-rotated state satisfies the same boundary conditions 
as the original state, 
\be
  e^{2\pi Ki\tl{X}}\brkt{U_\omg\left|n,\frac{j}{K}\right\rangle} 
 = \left|n,\frac{j}{K}\right\rangle, \;\;\;\;\;
 e^{i\tl{P}}\brkt{U_\omg\left|n,\frac{j}{K}\right\rangle} 
 = \left|n,\frac{j}{K}\right\rangle. 
 \label{Z3:periodic_cond}
\ee
Note that the left-hand-sides are rewritten as
\bea
 e^{2\pi Ki\tl{X}}\brkt{U_\omg\left|n,\frac{j}{K}\right\rangle} 
 \eql e^{-2\pi i(\phi_1-\phi_2)}U_\omg
 \left|n,\frac{j}{K}\right\rangle, \nonumber\\
 e^{i\tl{P}}\brkt{U_\omg\left|n,\frac{j}{K}\right\rangle} 
 \eql e^{2\pi i(\phi_1+2\phi_2-K/2)} 
 U_\omg\left|n,\frac{j}{K}\right\rangle, 
\eea
by using 
\be
 e^{i(2\pi K\tl{X}-\tl{P})} = e^{-\pi Ki}e^{2\pi K i\tl{X}}e^{-i\tl{P}}. 
\ee
Therefore, (\ref{Z3:periodic_cond}) indicates that
\be
 \phi_1-\phi_2 = 0, \;\;\;\;\;
 \phi_1+2\phi_2-\frac{K}{2} = 0, 
\ee
modulo 1. 
Namely, the allowed SS phases are 
\be
 \phi = \phi_1 = \phi_2 = \begin{cases} \displaystyle 
 0,\;\frac{1}{3},\;\frac{2}{3} & (\mbox{$K$: even}) \\[4mm] \displaystyle
 \frac{1}{6}, \; \frac{1}{2}, \; \frac{5}{6} & (\mbox{$K$: odd}) 
 \end{cases}. 
 \label{def:phi}
\ee
\subsubsection{Representation matrix for $\bdm{Z_3}$-rotation}
Notice that $U_\omg\left|n,\frac{j}{K}\right\rangle$ is not an eigenstate 
of $\tl{X}$ because
\be
 e^{2\pi i\tl{X}}\brkt{U_\omg\left|n,\frac{j}{K}\right\rangle} 
 = U_\omg\left|n,\frac{j+1}{K}\right\rangle. 
 \label{tlXofU_omg}
\ee
Thus it is expressed as a linear combination of 
the unrotated states: 
\be
 U_\omg\left|n,\frac{j}{K}\right\rangle 
 = \sum_j\cM_{jk}^{(\omg)}\left|n,\frac{j}{K}\right\rangle, 
\ee
where $\cM_{jk}^{(\omg)}$ are constants.\footnote{
Since $U_\omg$ commutes with $H$ in (\ref{def:H}), 
the ``energy eigenvalue''~$n$ does not change by 
the $Z_3$ rotation. 
} 
By summing (\ref{tlXofU_omg}) over $j$, we can identify it as
\be
 \sum_{j=0}^{\abs{K}-1}U_\omg\left|n,\frac{j}{K}\right\rangle 
 = \cN|n,0\rangle, 
\ee
where $\cN$ is a complex constant. 
By operating the ``shift operator''~$e^{-il\tl{P}/K}$ on both sides, 
we obtain 
\be
 \exp\brc{\frac{\pi i}{K}(-l^2-6l\phi)}
 \sum_{\tl{j}}\exp\brkt{-\frac{2\pi l\tl{j}i}{K}}U_\omg
 \left|n,\frac{\tl{j}}{K}\right\rangle 
 = \cN\left|n,\frac{l}{K}\right\rangle, 
\ee
where $\tl{j}\equiv j-l$. 
We have used (\ref{UXU:2}) with (\ref{def:phi}). 
Thus, it follows that
\bea
 \cN\sum_{l=0}^{\abs{K}-1}\exp\brc{\frac{\pi i}{K}(l^2+6l\phi)}
 \left|n,\frac{l}{K}\right\rangle 
 \eql \sum_{\tl{j}}\brc{\sum_{l=0}^{\abs{K}-1}
 \exp\brkt{-\frac{2\pi l\tl{j}i}{K}}}U_\omg
 \left|n,\tl{j}\right\rangle \nonumber\\
 \eql \abs{K}U_\omg|n,0\rangle, 
 \label{formula:1}
\eea
where we have used that
\be
 \sum_{l=0}^{\abs{K}-1}
 \exp\brkt{-\frac{2\pi l\tl{j}i}{K}} = \abs{K}\dlt_{\tl{j},0}. 
 \label{formula:2}
\ee
Using (\ref{formula:1}), we can show that
\bea
 U_\omg\left|n,\frac{j}{K}\right\rangle \eql 
 U_\omg e^{-ij\tl{P}/K}U_\omg^{-1}\cdot U_\omg|n,0\rangle 
 = e^{2\pi ji\tl{X}}U_\omg|n,0\rangle \nonumber\\
 \eql \frac{\cN}{\abs{K}}\sum_{l=0}^{\abs{K}-1}\exp\brc{
 \frac{\pi i}{K}(l^2+6l\phi+2jl)}\left|n,\frac{j}{K}\right\rangle, 
 \label{U_omgket}
\eea
and
\bea
 U_\omg^2\left|n,\frac{j}{K}\right\rangle \eql 
 \frac{\cN}{\abs{K}}\sum_l\exp\brc{
 \frac{\pi i}{K}(l^2+6l\phi+2jl)} \nonumber\\
 &&\times\frac{\cN}{\abs{K}}\sum_m\exp\brc{
 \frac{\pi i}{K}(m^2+6m\phi+2lm)}\left|n,\frac{m}{K}\right\rangle \nonumber\\
 \eql \frac{\cN^2}{\abs{K}^2}\sum_{\tl{l},m}
 \exp\brc{\frac{\pi i}{K}\brkt{\tl{l}^2+6\tl{l}\phi
 +2j(\tl{l}-m)}}\left|n,\frac{m}{K}\right\rangle \nonumber\\
 \eql \frac{\cN^2}{\abs{K}^{3/2}}e^{-\pi i(j+3\phi)^2/K+\pi i/4}
 \sum_m\exp\brkt{-\frac{2\pi jmi}{K}}
 \left|n,\frac{m}{K}\right\rangle, 
 \label{U_omg^2ket}
\eea
where $\tl{l}\equiv l+m$. 
At the last step, we used that
\bea
 \sum_{\tl{l}}\exp\brc{\frac{\pi i}{K}(\tl{l}^2+2j\tl{l}+6\tl{l}\phi)} 
 \eql \sum_{\tl{l}}\exp\brc{\frac{\pi i}{K}\brc{
 (\tl{l}+j+3\phi)^2-(j+3\phi)^2}} \nonumber\\
 \eql e^{-\pi i(j+3\phi)^2/K}\cdot\sqrt{\abs{K}}e^{\pi i/4}, 
\eea
which follows from the formula:
\be
 \sum_{s=0}^{M-1}\exp\brc{\frac{\pi i}{M}(s+t\pm\bt)^2} 
 = \sqrt{M}e^{\pi i/4} \;\;\;
 \mbox{for $t\in\mathbb{Z}$}, \;\;
 \bt = \begin{cases} 0 & (\mbox{for $M$: even}) \\[2mm] \displaystyle
 \frac{1}{2} & (\mbox{for $M$: odd}) \end{cases}. 
\ee
From (\ref{U_omg^2ket}), we obtain
\bea
 U_\omg^3|n,0\rangle \eql \frac{\cN^2}{\abs{K}^{3/2}}
 \exp\brc{\pi i\brkt{\frac{1}{4}-\frac{9\phi^2}{K}}}
 \sum_m U_\omg \left|n,\frac{m}{K}\right\rangle \nonumber\\
 \eql \frac{\cN^3}{\abs{K}^{5/2}}
 \exp\brc{\pi i\brkt{\frac{1}{4}-\frac{9\phi^2}{K}}} \nonumber\\
 &&\times\sum_l\exp\brc{\frac{\pi i}{K}(l^2+6l\phi)}
 \brkt{\sum_m\exp\brkt{\frac{2\pi mli}{K}}}
 \left|n,\frac{l}{K}\right\rangle \nonumber\\
 \eql \frac{\cN^3}{\abs{K}^{3/2}}\exp\brc{
 \pi i\brkt{\frac{1}{4}-\frac{9\phi^2}{K}}}|n,0\rangle. 
\eea
We have used (\ref{formula:2}). 
This should be $|n,0\rangle$ because $U_\omg^3=\id$. 
Therefore, the constant~$\cN$ is identified as
\be
 \cN = \sqrt{\abs{K}}\exp\brc{\pi i\brkt{
 -\frac{1}{12}+\frac{3\phi^2}{K}}}. 
\ee
Substituting this into (\ref{U_omgket}) and (\ref{U_omg^2ket}), 
we obtain 
\bea
 U_\omg \left|n,\frac{j}{K}\right\rangle 
 \eql \frac{1}{\sqrt{\abs{K}}}\exp\brc{\pi i
 \brkt{-\frac{1}{12}+\frac{3\phi^2}{K}}} \nonumber\\
 &&\times\sum_l\exp\brc{\frac{\pi i}{K}
 (l^2+6l\phi+2jl)}\left|n,\frac{l}{K}\right\rangle, \nonumber\\
 U_\omg^2 \left|n,\frac{j}{K}\right\rangle
 \eql \frac{1}{\sqrt{\abs{K}}}\exp\brc{\pi i
 \brkt{\frac{1}{12}-\frac{3\phi^2}{K}-\frac{j(j+6\phi)}{K}}}
 \nonumber\\
 &&\times \sum_l e^{-2\pi jl i/K}\left|n,\frac{l}{K}\right\rangle. 
\eea
From these expressions, we can read off the coefficients~$\cM_{jk}^{(\omg^l)}$ 
in (\ref{def:C_jk}) as (\ref{expr:C_jk:Z3}). 
\subsubsection{Mode functions}
The ``ground state''~$\left|0,\frac{j}{K}\right\rangle$ satisfies 
\be
 \hat{a}\left|0,\frac{j}{K}\right\rangle = 0, \label{cond:GS}
\ee
where
\bea
 \hat{a} \eql \frac{1}{2\sqrt{2\abs{c}}}\brkt{-\cP_2+i\cP_1} \nonumber\\
 \eql \frac{1}{\sqrt{2\abs{c}}}\brkt{\der_{\bar{z}}
 +\frac{\pi K}{2\Im\tau}z} = \sqrt{\frac{\Im\tau}{\pi\abs{K}}}D_{\bar{z}}. 
\eea
Thus, the solution of (\ref{cond:GS}) on $T^2$ is (\ref{def:cF}). 
}

\section{Magnitude of Yukawa coupling constants} \label{num_values}
\subsection{Three-generation case} \label{3genYukawa}
Here we collect numerical values of the Yukawa coupling constants in (\ref{Yukawa:3gen}). 
The eigenvalues of the matrices~$y^{(k)F}$ ($F=D,U,E,N$) are denoted by $\lmd^{(k)F}$. 
Their absolute values are calculated as
\bea
 |\lmd^{(1)D}| \eql |\lmd^{(1)E}| = \brkt{0.845,0.274,0.057}, \nonumber\\
 |\lmd^{(2)D}| \eql |\lmd^{(2)E}| = \brkt{0.921,0.350,0.321}, \nonumber\\
 |\lmd^{(3)D}| \eql |\lmd^{(3)E}| = \brkt{0.821,0.517,0.358}, \nonumber\\
 |\lmd^{(4)D}| \eql |\lmd^{(4)E}| = \brkt{0.644,0.524,0.208}, \nonumber\\
 |\lmd^{(5)D}| \eql |\lmd^{(5)E}| = \brkt{0.799,0.259,0.155}, \nonumber\\
 |\lmd^{(1)U}| \eql |\lmd^{(1)N}| = \brkt{0.731,0.279,0.0644}, \nonumber\\
 |\lmd^{(2)U}| \eql |\lmd^{(2)N}| = \brkt{0.921,0.350,0.321}, \nonumber\\
 |\lmd^{(3)U}| \eql |\lmd^{(3)N}| = \brkt{0.665,0.579,0.394}, \nonumber\\
 |\lmd^{(4)U}| \eql |\lmd^{(4)N}| = \brkt{0.769,0.415,0.220}, \nonumber\\
 |\lmd^{(5)U}| \eql |\lmd^{(5)N}| = \brkt{0.945,0.315,0.108}. 
 \label{lmd:Yukawa}
\eea

\subsection{One-Higgs-doublet case}
Here we collect numerical values of the Yukawa coupling constants 
in Sec.~\ref{1HD}. 
\subsubsection{$\bdm{\kp'=0}$ case}
The possible values of $n$, $\phi_1$ and $\phi_2$ in (\ref{expr:Y^pm}) are
\bea
 n \eql 0, \; 1, \; 2, \;\;\;\;\; \mbox{(mod 3)} \nonumber\\
 \phi_1 \eql \phi_2-{\rm floor}(\phi_2)+u, \;\;\;\;\; \mbox{(mod 4)} \nonumber\\
 \phi_2 \eql 0, \; \frac{1}{3}, \; \frac{2}{3}, \; 1, \; \frac{4}{3}, \; \frac{5}{3}, 
 \;\;\;\;\; \mbox{(mod 2)}
\eea
where $u=0,1,2,3$. 
For these values, only one generation is realized for each component. 
The absolute values of $Y^{(+)}(n,\phi_1,\phi_2)$ are listed 
in Table~\ref{k0case}. 
\begin{table}[t]
\begin{center}
\begin{tabular}{|c||c|c|c|c|c|c|}
 \hline \rule[-2mm]{0mm}{7mm} \backslashbox{$(n,u)$}{$\phi_2$} & 0 & 1/3 & 2/3 & 1
 & 4/3 & 5/3 \\ \hline\hline
 (0,0) & 0.573 &  &  & 0.191 &  &  \\ \hline
 (0,1) & 0.369 &  &  & 0.369 &  &  \\ \hline
 (0,2) & 0.191 &  &  & 0.573 &  &  \\ \hline
 (0,3) & 0.369 &  &  & 0.369 &  &  \\ \hline\hline  
 (1,0) &  & 0.522 & 0.270 &  & 0.522 & 0.811 \\ \hline
 (1,1) &  & 0.270 & 0.522 &  & 0.811 & 0.522 \\ \hline
 (1,2) &  & 0.522 & 0.811 &  & 0.522 & 0.270 \\ \hline
 (1,3) &  & 0.811 & 0.522 &  & 0.270 & 0.522 \\ \hline\hline 
 (2,0) & 0.811 & 0.369 & 0.191 & 0.270 & 0.369 & 0.573 \\ \hline
 (2,1) & 0.522 & 0.191 & 0.369 & 0.522 & 0.573 & 0.369 \\ \hline
 (2,2) & 0.270 & 0.369 & 0.573 & 0.811 & 0.369 & 0.191 \\ \hline
 (2,3) & 0.522 & 0.573 & 0.369 & 0.522 & 0.191 & 0.369 \\ \hline  
\end{tabular}
\end{center}
\caption{
The absolute values of $Y^{(+)}(n,\phi_1,\phi_2)$. 
The blanks denote cases in which the left- or the right-handed components do not have 
zero-modes. 
} \label{k0case}
\end{table}
Those of $Y^{(-)}$ are related to $\abs{Y^{(+)}}$ through 
\be
 \abs{Y^{(-)}(n,\phi_1,\phi_2)} 
 = \abs{Y^{(+)}(-n+2,\phi_1,\phi_2)}. 
\ee

\ignore{
\begin{description}
\item[$\bdm{n=0}$] \mbox{}\\
\begin{center}
\begin{tabular}{|c||c|c|c|c|c|c|}
 \hline \rule[-2mm]{0mm}{7mm} \backslashbox{$l$}{$\phi_2$} & 0 & 1/3 & 2/3 & 1
 & 4/3 & 5/3 \\ \hline\hline
 0 & 1.002 & 0.456 & 0.236 &  &  &  \\ \hline
 1 & 0.645 & 0.236 & 0.456 &  &  &  \\ \hline
 2 & 0.334 & 0.456 & 0.709 &  &  &  \\ \hline
 3 & 0.645 & 0.709 & 0.456 &  &  &  \\ \hline  
\end{tabular}
\end{center}
\item[$\bdm{n=1}$]
\begin{center}
\begin{tabular}{|c||c|c|c|c|c|c|}
 \hline \rule[-2mm]{0mm}{7mm} \backslashbox{$l$}{$\phi_2$} & 0 & 1/3 & 2/3 & 1
 & 4/3 & 5/3 \\ \hline\hline
 0 &  & 0.645 & 0.334 &  & 1.002 & 0.645 \\ \hline
 1 &  & 0.334 & 0.645 &  & 0.645 & 0.334 \\ \hline
 2 &  & 0.645 & 1.002 &  & 0.334 & 0.645 \\ \hline
 3 &  & 1.002 & 0.645 &  & 0.645 & 1.002 \\ \hline  
\end{tabular}
\end{center}
\item[$\bdm{n=2}$]
\begin{center}
\begin{tabular}{|c||c|c|c|c|c|c|}
 \hline \rule[-2mm]{0mm}{7mm} \backslashbox{$l$}{$\phi_2$} & 0 & 1/3 & 2/3 & 1
 & 4/3 & 5/3 \\ \hline\hline
 0 & 0.708 &  &  & 0.456 &  &  \\ \hline
 1 & 0.456 &  &  & 0.709 &  &  \\ \hline
 2 & 0.236 &  &  & 0.456 &  &  \\ \hline
 3 & 0.456 &  &  & 0.236 &  &  \\ \hline  
\end{tabular}
\end{center}
\end{description}
}

\subsubsection{$\bdm{\kp'=-1}$ case}
The absolute values of $y^{D,E}$ can be read off from Table.~\ref{k0case}. 
Those of $y^U(n_2,l,l')$ ($l=0,1$) are shown in Table.~\ref{k1case:U}. 
When $n_2=0$, $Q'_L$ does not have a zero-mode. 
\begin{table}[t]
\begin{center}
\begin{tabular}{|c||c|c|c|c|c|c|c|c|}
 \hline \rule[-2mm]{0mm}{7mm} \backslashbox{$(n_2,l)$}{$l'$} & 1 & 3 & 5 & 7
 & 9 & 11 & 13 & 15 \\ \hline\hline
 (1,0) &  & 0.365 &  &  & 0.667 &  &  & 0.461 \\ \hline
 (1,1) &  &  & 0.430 & &  & 0.798 &  &  \\ \hline
 (2,0) & 0.667 &  & 0.461 & 0.365 &  & 0.667 & 0.365 &  \\ \hline
 (2,1) & 0.798 & 0.430 &  & 0.430 & 0.430 & & 0.430 & 0.430  \\ \hline
\end{tabular}
\end{center}
\begin{center}
\begin{tabular}{|c||c|c|c|c|c|c|c|}
 \hline \rule[-2mm]{0mm}{7mm} \backslashbox{$(n_2,l)$}{$l'$} & 17 & 19 & 21 & 23
 & 25 & 27 & 29   \\ \hline\hline
 (1,0) & & & 0.667 &  &  & 0.365 &   \\ \hline
 (1,1) & 0.430 & & & 0.430 &  & & 0.430  \\ \hline
 (2,0) & 0.365 & 0.667 & & 0.365 & 0.461 & & 0.667  \\ \hline
 (2,1) & & 0.430 & 0.798 & & 0.430 & 0.430 &  \\ \hline
\end{tabular}
\end{center}
\caption{
The absolute values of $y^U(n_2,l,l')$. 
The blanks denote cases in which the left- or the right-handed components do not have 
zero-modes. 
} \label{k1case:U}
\end{table}
The coupling constants for the other values of $l$ are related 
to those in Table.~\ref{k1case:U} by
\bea
 \abs{y^U(n_2,2u,l')} \eql \abs{y^U(n_2,0,l'+8u)}, \nonumber\\
 \abs{y^U(n_2,2u+1,l')} \eql \abs{y^U(n_2,1,l'+8u)}. 
\eea
where $u=0,1,2,\cdots,14$.\footnote{
Note that $l$ and $l'$ are defined modulo 30. 
} 

The absolute values of $y^N(n_4,l,l')$ ($l=0,1$) are shown in Table.~\ref{k1case:N}. 
\begin{table}[t]
\begin{center}
\begin{tabular}{|c||c|c|c|c|c|c|c|c|c|}
 \hline \rule[-2mm]{0mm}{7mm} \backslashbox{$(n_4,l)$}{$l'$} & 1 & 3 & 5 & 7
 & 9 & 11 & 13 & 15 & 17\\ \hline\hline
 (0,0) & 0.559 & & 0.101 & 0.176 & & 0.176 & 0.101 & & 0.559  \\ \hline
 (0,1) & & 0.541 & 0.188 & & 0.288 & 0.288 & & 0.188 & 0.541 \\ \hline
 (1,0) & 0.559 & 0.533 & 0.101 & 0.176 & 0.533 & 0.176 & 0.101 & 0.533 & 0.559 \\ \hline
 (1,1) & 0.924 & 0.541 & 0.188 & 0.924 & 0.288 & 0.288 & 0.924 & 0.188 & 0.541 \\ \hline
 (2,0) & 0.559 &  & 0.101 & 0.176 &  & 0.176 & 0.101 & & 0.559 \\ \hline
 (2,1) & & 0.541 & 0.188 & & 0.288 & 0.288 &  & 0.188 & 0.541 \\ \hline
\end{tabular}
\end{center}
\caption{
The absolute values of $y^N(n_4,l,l')$. 
The blanks denote cases in which the left- or the right-handed components do not have 
zero-modes. 
} \label{k1case:N}
\end{table}
The coupling constant for the other values of $l$ are related to 
those in Table.~\ref{k1case:N} by
\bea
 \abs{y^N(n_4,2u,l')} \eql \abs{y^N(n_4,0,l'-2u)}, \nonumber\\
 \abs{y^N(n_4,2u+1,l')} \eql \abs{y^N(n_4,1,l'-2u)}. 
\eea
where $u=0,1,2,\cdots,8$.\footnote{
Note that $l$ and $l'$ are defined modulo 18. 
}


\end{document}